# Impact of Cu–Mn ratio on Structure and Defects in Layered Multiferroic $Cu_{1-x}Mn_{1+y}SiTe_3$

Sai Venkata Gayathri Ayyagari[1], Boyang Zheng[2,3], Sreekant Anil[1], Subrata Ghosh[2,3], Yuxi Zhang[1], Yu Liu[2,3], Chandan De[2,3], Ke Wang[4], Jeffrey Shallenberger[4], Weiwei Xie[5], Vincent H. Crespi[2,3], Zhiqiang Mao[1,2,3], and Nasim Alem[1*]

[1]Department of Materials Science and Engineering, The Pennsylvania State University, University Park, PA 16802, USA

[2]2D Crystal Consortium, Materials Research Institute, The Pennsylvania State University, University Park, PA 16802, USA

[3]Department of Physics, The Pennsylvania State University, University Park, PA 16802, USA

[4]Materials Research Institute, University Park, PA 16802, USA

[5]Department of Chemistry, Michigan State University, East Lansing, MI 48864, USA

* Corresponding author: nua10@psu.edu

**Abstract**

Multiferroic materials exhibit the coexistence of magnetic and ferroelectric order, enabling control of magnetism through electric fields and vice versa. These properties make them attractive for spintronic and memory device applications. Recent studies on $Cu_{1-x}Mn_{1+y}SiTe_3$ ($0.04 \leq x \leq 0.26$; $0.03 \leq y \leq 0.15$) have revealed strong magnetoelectric coupling, with variations in Mn-to-Cu concentration leading to variations in optical, electronic, and magnetic responses. Despite these findings, the influence of nanoscale structure and defects on the observed properties remains poorly understood. In this study, we investigate the structure and nanoscale defects in Cu-deficient $Cu_{1-x}Mn_{1+y}SiTe_3$ (Cu:Mn ratio <1, i.e., with $0.04 \leq x < 0.26$ and $0.03 \leq y \leq 0.15$) and Cu-rich $Cu_{1+x}Mn_{1-y}SiTe_3$ (Cu:Mn ratio >1, i.e., with $0.04 \leq x \leq 0.3$ and $0.13 \leq y \leq 0.31$) crystals using scanning/transmission electron microscopy and single-crystal X-ray diffraction. Cu-deficient crystals exhibit extensive stacking faults correlated with chemical inhomogeneity between Mn and Cu, along with variations in Te stacking. In contrast, Cu-rich crystals show fewer stacking faults but contain other local structural variations, such as needle-shaped precipitates and loop-like features. These distinct local structural features between Cu-rich and Cu-deficient crystals can be correlated to variations in their observed properties. Complementary density functional theory calculations confirm that the Cu-rich structure is more polar than the Cu-deficient structure. Overall, this study provides a comprehensive understanding of how subtle changes in chemistry influence the nanoscale structure, defect distribution, and functional properties in $Cu_{1-x}Mn_{1+y}SiTe_3$, offering guidance for designing multiferroic materials with tailored performance.

## 1. Introduction

Materials that combine seemingly incompatible phenomena within a single system offer a powerful route to realizing unusual functional properties. For example, thermoelectric materials are designed to exhibit high electrical conductivity together with low thermal conductivity, a combination that is counterintuitive in conventional solids. In an analogous way, the coexistence of ferroelectricity and magnetism is surprising because ferroelectricity is typically favored in systems with empty d shells, whereas magnetism requires partially filled d orbitals. Materials that exhibit the coexistence of more than one ferroic order, such as ferroelectricity, ferromagnetism, and ferroelasticity within the same phase, are known as multiferroic materials. The coupling between these ferroic orders gives rise to multifunctional behavior, making multiferroics attractive for applications in data storage, sensing, actuation, and spintronic devices [1–3]. Among known systems, $BiFeO_3$ has been investigated most extensively, while other compounds such as $LuFe_2O_4$, $YMnO_3$, and several others have also been well studied for their distinct ferroic coupling mechanisms [2,3].

Despite these discoveries, a key challenge limiting practical applications is the loss of ferroelectricity at nanoscale dimensions, motivating the exploration of layered materials that can retain these properties at reduced length scales [4,5]. To address these challenges, significant efforts have been devoted to the search for layered multiferroics, leading to the discovery of several layered systems [6,7]. Among them, $Cu_{1-x}Mn_{1+y}SiTe_3$ ($0.04 \leq x \leq 0.26$; $0.03 \leq y \leq 0.15$) has recently emerged as a particularly interesting candidate. This material exhibits room-temperature ferroelectricity, canted antiferromagnetic ordering below $T_N$=35 K, and strong magnetoelectric coupling, with a magnetically induced polarization of 0.8 μC/cm² [8]. This unique combination of properties establishes $Cu_{1-x}Mn_{1+y}SiTe_3$ as a compelling prototype for exploring emergent functionalities in layered multiferroic systems.

Recent studies further show that stoichiometric variation in this system strongly influences its electric polarization. In particular, Cu-rich crystals with Cu:Mn > 1 ($Cu_{1+x}Mn_{1-y}SiTe_3$ ($0.04 \leq x \leq 0.3$; $0.13 \leq y \leq 0.31$)) exhibit an enhanced second-harmonic generation (SHG) coefficient compared with Cu-deficient crystals with Cu:Mn < 1 [9]. Although Cu-deficient and Cu-rich crystals have similar band gaps, Cu-rich crystals show doped semiconducting behavior with enhanced carrier density, whereas Cu-deficient crystals are insulating below $T_N$ [8]. To understand the origin of these stoichiometry-dependent property changes, it is therefore essential to examine the nanoscale structure and defect landscape of Cu-deficient and Cu-rich crystals and to determine how these structural features govern the observed macroscopic properties.

Defects are inherently present in all materials, and often strongly affect the performance of ferroic materials. Understanding the role of defects in ferroelectrics is crucial for practical applications, as dislocations and point defects can pin domain walls and modify the shape of the P-E loop [10,11]. Stacking faults have been linked to dielectric anomalies arising from local electrical polarizations, where inversion symmetry is broken by in-homogeneously distributed faults; for instance, in α-$RuCl_3$ such faults have been associated with multiple magnetic transitions [12]. Furthermore, stacking faults can induce local symmetry breaking in centrosymmetric structures, giving rise to ferroelectricity in layered materials [13]. These findings emphasize the importance of investigating the local structure and defect characteristics in the newly discovered $Cu_{1-x}Mn_{1+y}SiTe_3$ system to elucidate how structural and chemical variations influence its physical properties.

In this work, we explore the chemical and atomic structure of $Cu_{1-x}Mn_{1+y}SiTe_3$ as well as the structure of its defects from Cu-deficient to Cu-rich compositions using scanning/transmission electron microscopy (S/TEM) and single-crystal X-ray diffraction combined with density functional theory (DFT) calculations. We find that the Cu-deficient phase exhibits *Pm* symmetry characterized by a high density of stacking faults accompanied by an increase in local Cu content and a decrease in Mn content. In contrast, while the Cu-rich phase maintains similar *Pm* symmetry, it displays a significantly lower stacking fault density whose thickness is of only a few Te layers. Furthermore, the Cu-rich phase hosts unique needle-shaped features with even higher local Cu concentrations. DFT calculations indicate that the stacking fault regions are energetically less stable than the pristine lattice in the Cu-deficient crystal. Furthermore, the DFT calculations suggest that the Cu-rich crystal exhibits *p*-type doping and enhanced polarity, which is consistent with our experimental observations. Overall, this work aims to uncover the relationships among composition, local structure, and defect chemistry in $Cu_{1-x}Mn_{1+y}SiTe_3$, providing a foundation for understanding and tailoring functional properties in emerging multiferroic systems.

## 2. Results and Discussion

The composition of synthesized $Cu_{1-x}Mn_{1+y}SiTe_3$ crystals varies based on the nominal composition of the starting materials and the growth conditions. Samples characterized by Cu deficiency and Mn enrichment (i.e., with $0.04 \leq x < 0.26$ and $0.03 \leq y \leq 0.15$, Cu:Mn ratio <1 ) [8] are referred to as Cu-deficient. Conversely, crystals with higher Cu content (i.e., with $-0.30 \leq x < -0.04$ and $-0.31 \leq y \leq -0.13$, Cu:Mn ratio > 1) are referred to as Cu-rich crystals. In this paper, Cu-rich crystals are represented as $Cu_{1+x}Mn_{1-y}SiTe_3$ with $0.04 \leq x \leq 0.3$ and $0.13 \leq y \leq 0.31$.

To determine the average crystal structures and their structural evolution from Cu-deficient to Cu-rich samples, single-crystal X-ray diffraction measurements were performed (Figure S1-S3). Structural analysis reveals that both compositions crystallize in a monoclinic, non-centrosymmetric space group *Pm*. A detailed refinement of site occupancies shows that in the Cu-deficient crystal ($Cu_{0.79}Mn_{1.08}SiTe_3$), excess Mn partially substitutes for Cu on the Cu sites, leading to mixed occupancy. In contrast, the Cu-rich crystals exhibit additional occupied atomic sites relative to the Cu-deficient phase. These results highlight how variations in Cu content influence both the site occupancy and the local structure without altering the average fundamental space group symmetry.

### *2.1 Local Structural Investigation of Cu-deficient $Cu_{1-x}Mn_{1+y}SiTe_3$ ($0.04 \leq x \leq 0.26$ and $0.03 \leq y \leq 0.15$)*

To investigate the nanoscale structure and defects, we first examine the Cu-deficient sample with the composition ($Cu_{0.78}Mn_{1.10}SiTe_{3.02}$). We performed both real-space and reciprocal-space analysis utilizing selected-area electron diffraction (SAED) and scanning transmission electron microscopy (STEM) imaging. Figure 1a shows the SAED pattern along [010] zone, where sharp reflections (excluding streaks) are consistent with Pm symmetry, in agreement with previously reported single-crystal X-ray and neutron refinements [8,14]. It is well established that such streaks in electron diffraction denote deviations from an ideal three-dimensional crystal lattice, indicating the presence of defects such as stacking faults [15]. The streaking in diffraction pattern occurs along the c direction, suggesting that defects (stacking faults) are preferentially observed on the (001) plane and are apparent throughout the sample in the TEM images (Figure S4). Similar streaks along the c direction were also observed in the single crystal X-ray results (Figure S2).

To further examine the atomic arrangement and defects, atomic-resolution STEM imaging was performed. As the sample exhibits beam sensitivity (see Figure S5e and S5f), STEM imaging was conducted at an accelerating voltage of 80 kV. Figure 1b displays the ADF-STEM image, where the blue inset highlights a pristine region overlaid with the corresponding cif file from single crystal X-ray refinement, while the magenta region indicates an area containing defects. The yellow line connecting Te atoms illustrates the shift in Te stacking within the defect region, confirming the formation of stacking faults.

In the enlarged blue inset (Figure 1c), Te atoms, possessing higher atomic numbers (Z), appear distinctly larger, and the atoms located between the Te layers are overlaid with Cu, Mn, and Si positions. Owing to the projected interatomic distance between Mn and Si being less than 60 pm, their individual locations are difficult to distinguish. In Figure 1d, while Te atoms are well resolved, the intermediate layer between Te layers remains indistinct in the defect region. This limitation necessitated imaging along an alternative zone axis to elucidate the atomic arrangement in the stacking fault region (Figure S5d). Due to beam sensitivity, only variation in Te arrangement can be clearly imaged along [100] zone, thereby making it challenging to conclude the structure and how it may influence the overall properties of this multiferroic material.

To further investigate the chemical and electronic variations between stacking fault regions and pristine areas, spectroscopic analyses were carried out using both EDS and EELS techniques. The EDS maps presented in Figures 1e–g reveal local chemical changes associated with stacking fault regions. Examination of EDS results shows that stacking fault regions are enriched in Cu relative to Mn, while the pristine regions exhibit a more uniform intermixing of Mn and Cu. Figure 1h displays the summed intensity from each row of Mn and Cu in the EDS map, clearly illustrating the chemical inhomogeneity within the defect regions. EELS analysis (Figure S5) demonstrates that there is no variation in the Mn edge across the stacking fault and pristine regions, consistent with $Mn^{2+}$ throughout the sample. Strain maps using geometric phase analysis reveal pronounced strain fields at the stacking fault interfaces (Figure S6). The presence of high-density stacking faults likely suppresses the macroscopic polarization in the Cu-deficient sample, as evidenced by the reduced SHG intensity in our previous report[8]. Furthermore, these structural defects may be associated with the competition between the observed short range magnetic correlation and the long-range antiferromagnetic ordering, as discussed in our earlier work[8].

DFT calculations indicate that the Te stacking in the pristine region, referred to as ABC stacking, is energetically more stable than the Te stacking associated with the stacking-fault region, referred to as AB stacking (detailed discussion in SI, Section S7). This trend is consistent with STEM observations, indicating stacking-faults as the first regions degrading under electron-beam irradiation (Figure S5(e,f)).

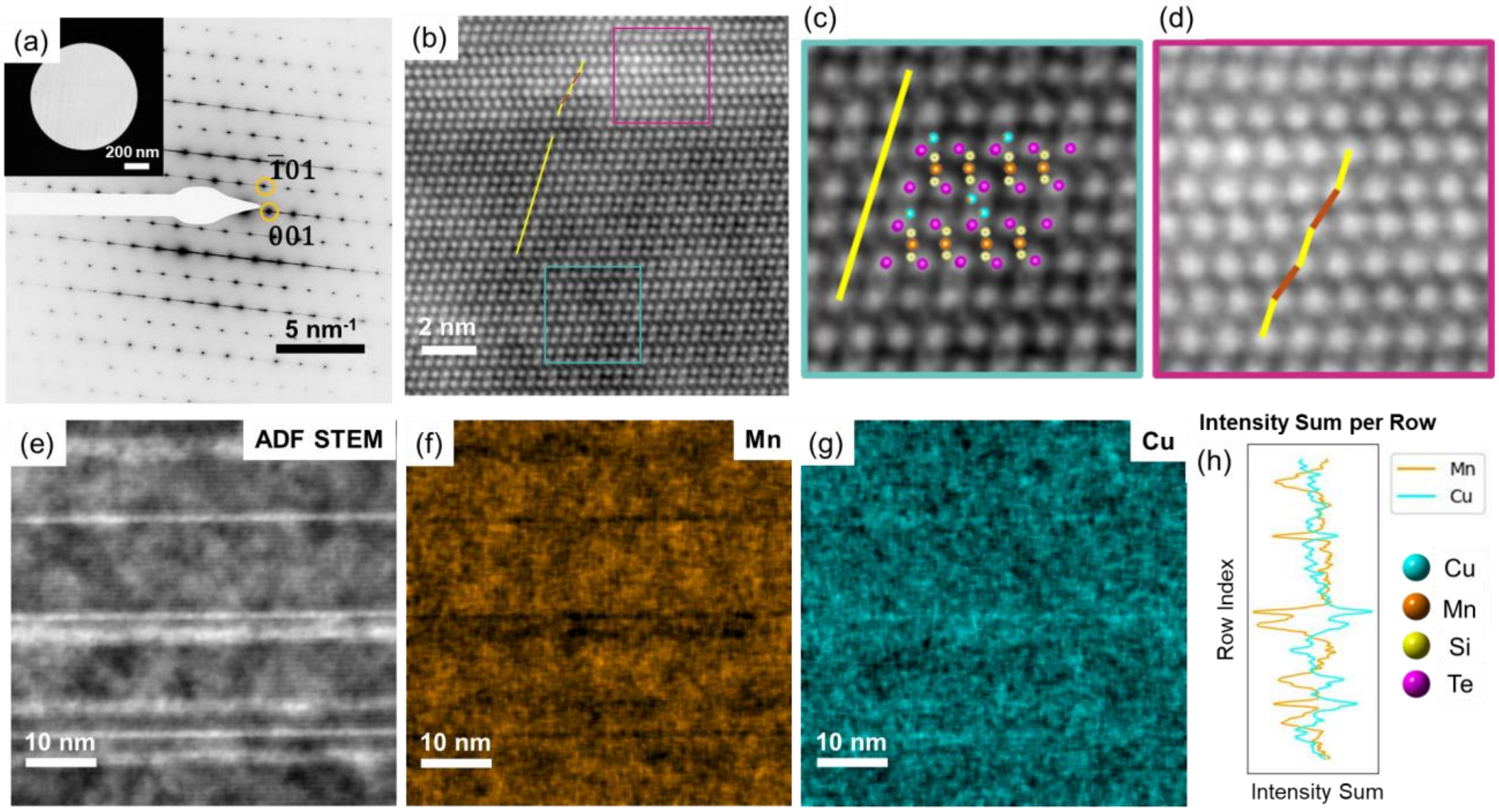

**Figure 1. Structural investigation of Cu-deficient crystal ($Cu_{0.78}Mn_{1.10}SiTe_{3.02}$) reveals stacking faults:** (a) SAED of $Cu_{0.78}Mn_{1.10}SiTe_{3.02}$ cross-section specimen along [010] zone axis, with an inset showing the selected area. (b) Atomic-resolution ADF-STEM image from [010] zone axis. (c) Magnified view of the pristine ADF-STEM image overlaid with the corresponding structure model. (d) Magnified image of the defect (stacking fault) region showing a change in Te stacking. (e-g) Elemental distribution maps of Mn and Cu from the ADF-STEM region shown in (e). (h) Intensity per sum row plot corresponding to (f) and (g).

### *2.2 Local Structural Investigation of Cu-rich $Cu_{1+x}Mn_{1-y}SiTe_3$ ($0.04 \leq x \leq 0.3$ and $0.13 \leq y \leq 0.31$)*

Advanced electron microscopy imaging and spectroscopy were performed to uncover the structure of Cu-rich crystal with the composition $Cu_{1.14}$ $Mn_{0.77}SiTe_{2.9}$. Figure 2 shows the selected area electron diffraction and STEM images in $Cu_{1.14}Mn_{0.77}SiTe_{2.9}$ crystal along [100] and [010] zone axis. Unlike Cu-deficient crystals, no streaking along the c direction is observed in the diffraction patterns indicating no stacking faults in the (001) planes. This observation is also consistent with the absence of streaks in X-ray reciprocal-space analysis (Figure S3). The electron diffraction along both zones agree with a monoclinic structure with Pm symmetry and the refined structure obtained from single crystal analysis[16]. The atomic structure of the crystal is shown in the ADF-STEM images (Figure 2c and 2f) along both zones with the structure model from single X-ray refinement overlaid on top.

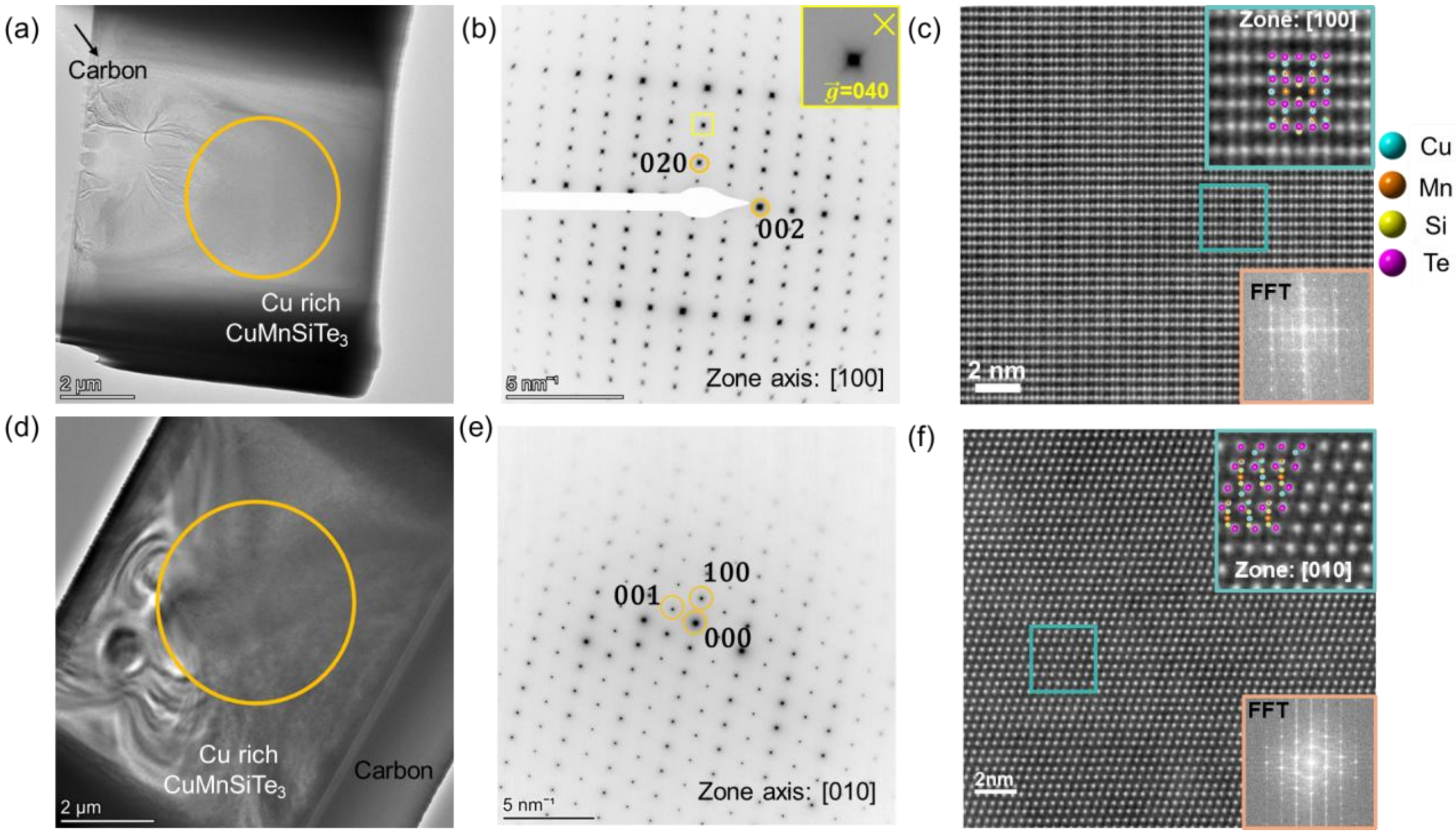

**Figure 2. Overall structure of Cu-rich ($Cu_{1.14}Mn_{0.77}SiTe_{2.9}$) crystals:** (a) TEM image of $Cu_{1.14}Mn_{0.77}SiTe_{2.9}$ cross-section specimen taken along the [100] zone axis. (b) SAED pattern from the yellow circled region in (a). The inset shows a magnified view of a reflection. (c) Atomic-resolution ADF-STEM image along the [100] zone axis, with insets showing corresponding FFT and magnified image overlaid with the structure model. (d) TEM image of $Cu_{1.14}Mn_{0.77}SiTe_{2.9}$ cross-section specimen taken along [010] zone axis. (e) SAED pattern from the yellow circled region in (d). (f) Atomic-resolution ADF-STEM image along the [010] zone axis. Diffuse rings in the [010] FFT arise from FIB-induced amorphization. We note that because Te is a heavier element and the interatomic spacings along [010] axis fall below the detection limit, it is difficult to resolve the atoms between the Te layers with high precision.

### *2.2.1 Needle-Shaped Features in Cu-rich crystals*

A closer look at the SAED along [100] zone shows cross-streaking in the reflections highlighted in the inset region (presented in Figure 2b) which are absent in the reflections along [010] zone. Figure 3a presents a TEM image with a higher magnification compared to Figure 2a, revealing directional needle-shaped features in the Cu-rich crystals along [100] zone axis. These features are reminiscent of precipitates reported in Al-Cu based alloys in the literature[17,18]. The SAED pattern in Figure 3b displays streaking in the reflections perpendicular to the features observed in the real space image in Figure 3a. Diffraction analysis identified that these features correspond to the {021} family of planes. The angle between these needle shaped features were measured on ranging between 102° to 120°, with an average angle of 111 $\pm$ 0.6° as shown in Figure 3e and Figure S8.

Elemental distribution maps (Figure 3e-g and Figure S9) reveal chemical inhomogeneity across the needle-shaped features in the Cu-rich $Cu_{1.14}Mn_{0.77}SiTe_{2.9}$ sample, with Cu enrichment and Mn deficiency localized to these directional defects, while the surrounding matrix is Mn-rich. These observations suggest the possibility of kinetic trapping of excess Cu during growth, leading to localized site occupancy changes akin to Cu-rich precipitates previously reported in the high entropy oxide (Mg,Co,Ni,Cu,Zn)O[19].

Atomic-resolution STEM imaging and unsupervised machine learning analysis (Figure 3c and Figure S10) further uncover subtle structural changes in the precipitates relative to the matrix. While the real-space atomic-resolution STEM image in Figure 3d shows no evident changes in Te atomic arrangements but visible contrast variations between the Cu rich (needle shaped feature) and Cu deficient (matrix) region, the corresponding FFT patterns (Figure S10) reveal structural differences between the precipitates and matrix. Series of STEM images acquired at different collection angles (camera length) on the same region, as shown in Figure S11 highlights that these needle-shaped features exhibit strong diffraction contrast rather than Z contrast. Strain maps (Figure S12) reveal tensile $\varepsilon_{xx}$ within the needle-shaped features alongside differential strain fields between precipitates and matrix, linking chemical inhomogeneity to elastic distortion.

Aside from the average structural result refined from X-ray data, DFT calculations suggest that the Cu-rich sample could also adopt the structures shown in Figure S13. In this scenario, the bright "needles" (Cu-rich region from EDS) correspond to the structure in Figure S13a, whereas the darker regions (matrix region, Cu-deficient from EDS) correspond to the structure in Figure S13b, which has one fewer Cu in the unit cell as compared to Figure S13a. The ratio for Te-layer distances (Si-dimer : Cu-layer) is 1.080 and 1.091 for the two suggested structures. These two numbers are closer to the value measured in TEM images (~1.08 in Figure S14), while the X-ray–refined structure gives a ratio of ~1.03. By counting valence electrons, these two structures proposed by DFT will be $p$-doped against the stoichiometric $MnCuSiTe_3$, consistent with the experimental finding that the Cu-rich sample is $p$-doped. These two structures have short Cu-Cu distances, and further experimental investigations sensitive to local chemical environments like NMR could help confirm their existence.

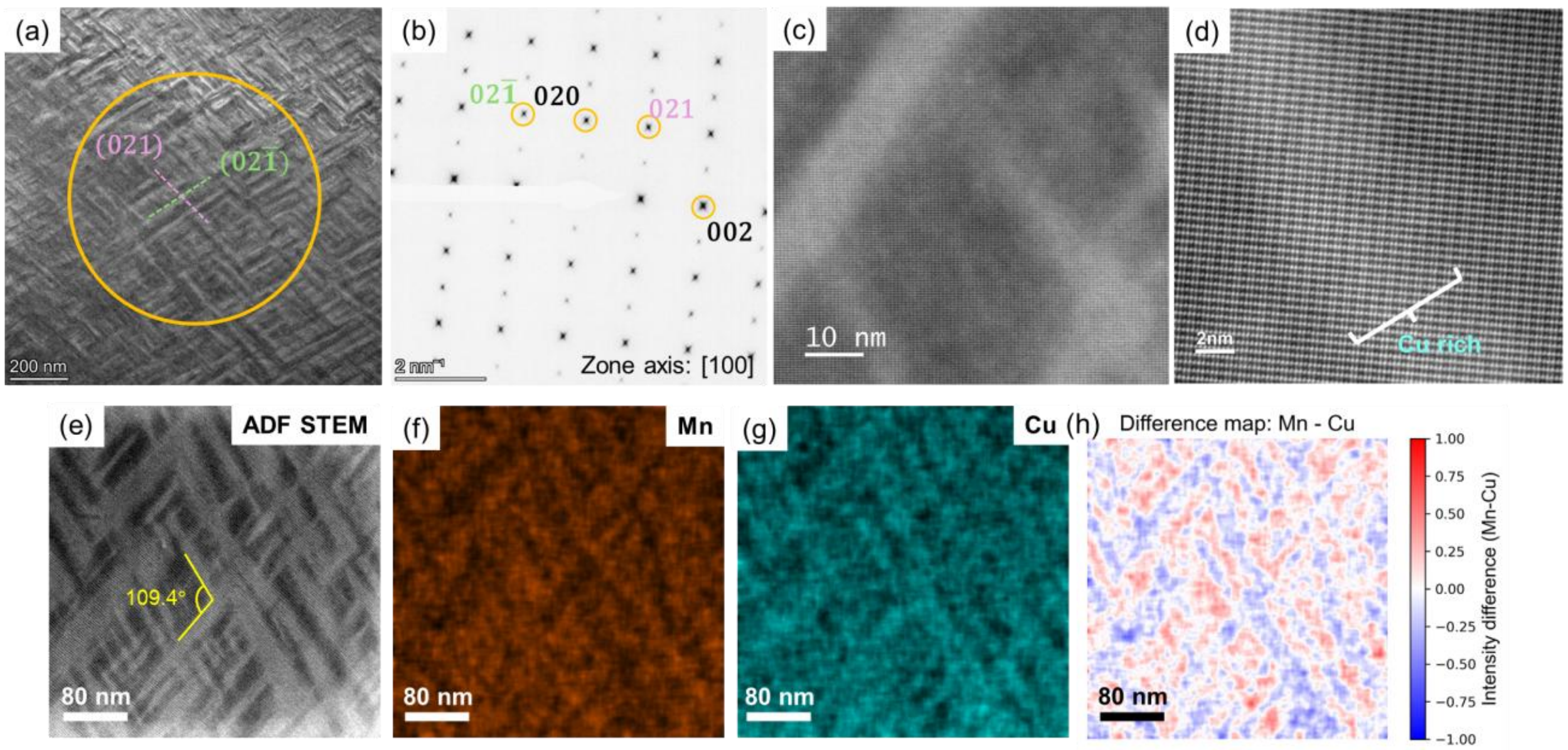

**Figure 3. Cu-rich needle-shaped features in Cu-rich ($Cu_{1.14}Mn_{0.77}SiTe_{2.9}$) crystals:** (a) TEM image of $Cu_{1.14}Mn_{0.77}SiTe_{2.9}$ cross-section specimen along [100] zone axis, highlighting the needle-shaped features. (b) SAED pattern acquired from the yellow circled region in (a). (c) ADF STEM image of the needle-shaped features, with a magnified view shown in (d). (e-g) ADF-STEM image and corresponding elemental distribution maps for Mn and Cu across the defect region. (h) Intensity difference map (Mn minus Cu) illustrating chemical inhomogeneity across these features.

*2.2.2 Stacking-Fault like Defect in Cu-rich crystals*

Further investigation of the Cu-rich crystals reveals other type of defects in the Cu-rich crystals along the [010] and [100] zones (Figure 4 and Figure S15). Figure 4a shows the stacking fault like defect observed in a small region of Cu-rich crystals (less than 410 nm wide) near the top, close to the carbon layer deposited during FIB sample preparation. Electron diffraction and imaging (Figure 4a and 4b) reveal streaking in both real space and reciprocal space, confirming the presence of stacking faults. ADF-STEM imaging indicates these faults involve a change in Te atomic arrangement, similar to that seen in Cu-deficient sample. Unlike the Cu-deficient crystals, where excessive stacking fault formation occurs across multiple Te layers, the stacking faults in Cu-rich crystals are less dense, sporadic, and observed only on a few Te layers.

Elemental maps (Figures 4e-g) confirm that defect regions are rich in Cu and deficient in Mn. Figure 4d reveals the magnified ADF STEM image of the defect (Cu rich) region with additional internal stacking fault lines within the defect region itself. The structure of this Cu-rich defect region is noticeably different from that of the pristine region highlighting that this stacking fault like defect is adapting a different crystal symmetry, corroborated by FFT changes (Figure S15) in this defect region. Strain maps (Figure S16) further reveal compressive $\varepsilon_{xy}$ along the stacking fault line, with distinct strain fields in the defect vs. pristine regions linking chemical segregation to structural changes.

Due to the very weak Mn signal in the Cu-rich defect region, we identify the $Cu_2SiTe_3$ (mp-675120) structure from the Materials Project database [20] as the most probable candidate for this region. This identification is supported by the close agreement between the database's Te-layer distance (3.42 Å) and the 3.44 Å value observed experimentally (Figure S17), as well as the alignment of atomic positions with STEM imaging (Figures 4d and S17). In this defect region of $Cu_2SiTe_3$, the structure would be polar; however, we do not expect these regions to be major contributors to the polarization observed in the sample, because the opposite polarity of different domains (Figure 4d) would effectively cancel out their individual contributions.

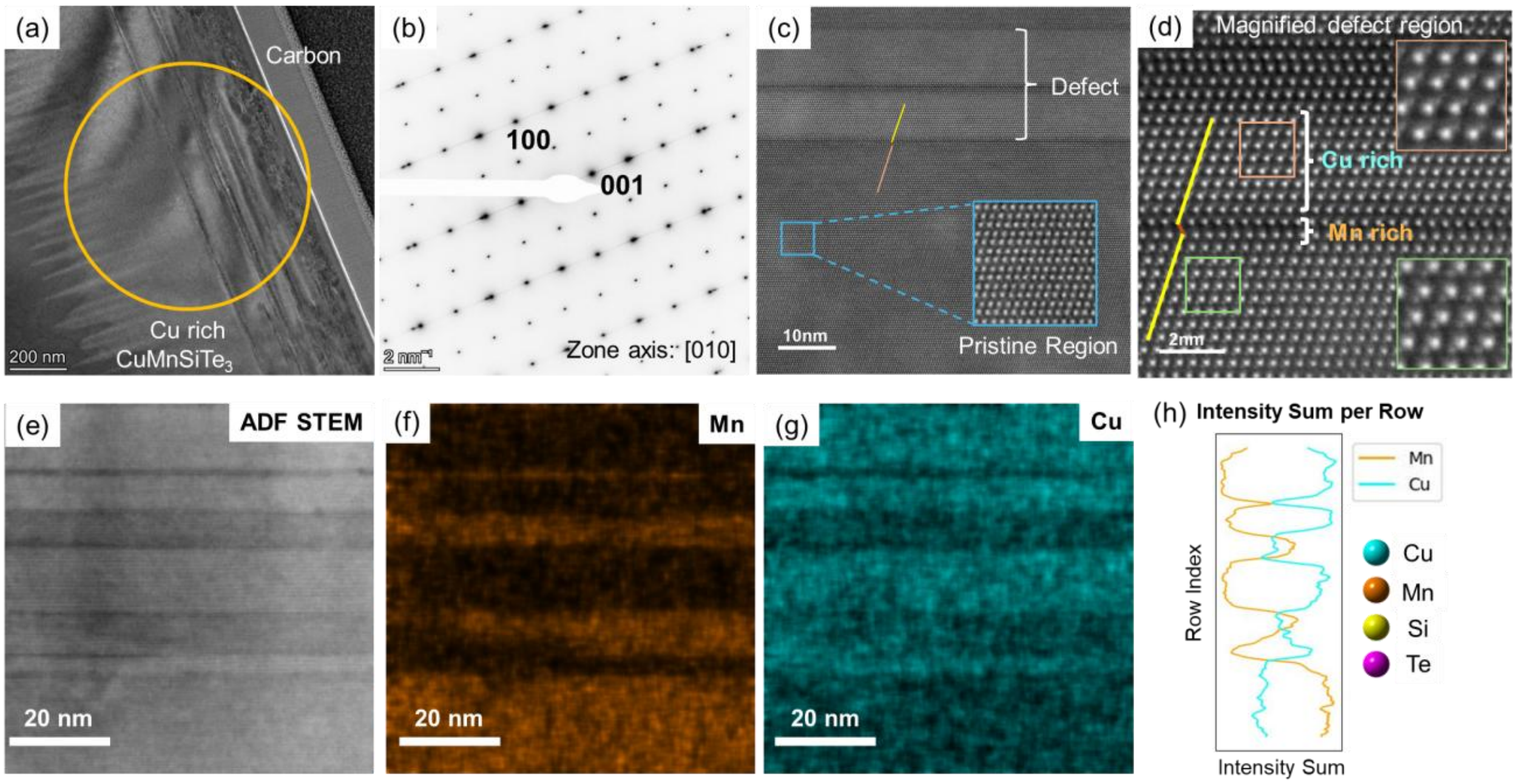

**Figure 4. Stacking-fault-like defect in Cu-rich crystals are rich in Cu and exhibit different structural arrangement:** (a) TEM image of $Cu_{1.14}Mn_{0.77}SiTe_{2.9}$ cross-section specimen along [010] zone axis. (b) SAED pattern from the yellow circled region in (a). (c) ADF-STEM image of the stacking fault region with inset showing the change in Te arrangement. (d) Magnified ADF-STEM image of the defect region showing changes in the Te arrangement and also showing a change in the atomic structure in the defect region compared to pristine region (e-g) ADF-STEM image and corresponding Mn and Cu elemental distribution maps across the stacking fault region. (h) Intensity per sum row plot corresponding to (f) and (g).

### *2.2.3 Loop-like Features in Cu-rich crystals*

Additionally, loop-like features are observed in the Cu-rich crystals when imaged along both [100] and [010] zone axis, as shown in Figure S19 and S20. These features do not show any detectable chemical inhomogeneity but exhibit strong diffraction contrast, likely arising from strain fields or from ferroelectric domain structures [21,22]. To confirm whether the observed loop-like features correspond to ferroelectric domains and to establish the nature of associated domain walls, further experiments such as testing to see Friedel's-pair-breaking contrast conditions using $\pm \vec{g}$, or performing in situ biasing experiments will be the subject of future investigations.

### *2.3 Polarization Estimate using DFT on Proposed Structures*

Interestingly, SHG measurements also indicate that the Cu-rich crystals exhibit higher SHG coefficient, i.e. polarization than the Cu-deficient crystals [9]. The polarization can be estimated from Born effective charges calculated from density functional theory (DFT) [23],

$$P = \frac{1}{V_{cell}} \sum Z^{*} e u$$

where, $V_{\mathrm{cell}}$ is the volume of the cell, $Z^{*}$ is the Born effective charge, $e$ is the elementary charge, $u$ is the displacement from the reference structure which is non-polar. Using the Cu-rich structure (details in the Methods section), the Mn in the metal layer has a Born effective charge of +1.06, and the Cu in the metal layer has an average value of +1.03. Referenced to centrosymmetric positions, i.e., the octahedron centers, the displacements in the $z$-direction are –0.86 Å for Mn, 0.68 and –0.73 Å for Cu. Plugging these numbers back into the equation with the corresponding composition, we get 1.1 eÅ/cell ~ 2.8 μC/cm$^2$ for the Cu-rich structure, and similarly, for alternative models for the Cu-rich structures, 0.48 eÅ/cell ~ 1.2 μC/cm$^2$ for Figure S13a and 0.042 eÅ/cell ~ 0.11 μC/cm$^2$ for Figure S13b, which averages to 0.26 eÅ/cell ~ 0.66 μC/cm$^2$. For the Cu-deficient structure, this method estimates to 0.065 eÅ/cell ~ 0.17 μC/cm$^2$. This estimation shows that the Cu-rich structure is more polar than the Cu-deficient structure, consistent with the experimental findings.

## 3. Conclusion

In this study, advanced electron microscopy imaging and spectroscopy analysis reveal that changes in the Cu: Mn content in $Cu_{1-x}Mn_{1+y}SiTe_3$ profoundly influence local defect landscapes, while preserving the monoclinic Pm symmetry. Cu-deficient crystals (Cu:Mn < 1) exhibit prevalent (001) stacking faults driven by local Cu-enrichment and accompanied by significant strain fields. In contrast, Cu-rich crystals (Cu:Mn > 1) develop needle-shaped Cu-precipitates, sporadic stacking-fault-like defects, and loop-like features. This demonstrates how subtle stoichiometric changes tune nanoscale structure across the series leading to profound changes in the electronic (p-doped semiconductor in case of Cu-rich vs. insulator (below $T_N$) in Cu-deficient) and optical (enhanced SHG in Cu-rich vs. Cu-deficient) responses.

These structural observations, combined with DFT, enables the estimate of polarization in the proposed structures. Our DFT calculations (polarization of ~2.8 $\mu C/cm^2$ in Cu-rich vs. ~0.17 $\mu C/cm^2$ in Cu-deficient) provide atomic-scale insight into the structural origins of this composition-dependent response. Our findings establish $Cu_{1-x}Mn_{1+y}SiTe_3$ as a composition-tunable platform for studying defect-driven properties in layered chalcogenides. These insights open pathways for engineering magnetoelectric functionality through precise nonstoichiometric control.

## 4. Methods

***Single Crystal Growth:***

The single crystals of $Cu_{1-x}Mn_{1+y}SiTe_3$ were grown using a melt-grown technique. The growth details have been previously reported by De *et al.*[8].

***Single Crystal X-ray Analysis:***

Single crystals were mounted on nylon loops using Paratone oil and measured at room temperature on a Rigaku XtaLAB Synergy Dualflex diffractometer equipped with a HyPix detector. Diffraction data were collected using ω scans with Mo Kα radiation (λ = 0.71073 Å) generated from a micro-focus sealed X-ray tube (50 kV, 1 mA). Data collection strategies were optimized using CrysAlisPro (v1.171.43.120a). Data reduction included Lorentz and polarization corrections, along with a numerical absorption correction based on Gaussian integration over a multifaceted crystal model [24]. An empirical absorption correction using spherical harmonics was applied with the SCALE3 ABSPACK algorithm [25]. The crystal structures were solved and refined using Olex2 in conjunction with the SHELXTL package [26–28].

***Scanning/Transmission Electron Microscopy:***

**Sample preparation:**

Thin lamellae for TEM analysis were prepared using a Thermo Fisher Helios NanoLab Dual Beam Focused Ion Beam system at Penn State. The lamellae were extracted at an accelerating voltage of 30 kV, thinned at 5 kV, and finally cleaned at 2 kV. The sample exhibited sensitivity to beam conditions; therefore, most of the thinning process was performed at 5 kV to minimize ion beam–induced damage. Laue diffraction was performed on the crystals to identify the crystal orientation, and TEM analyses were conducted along the [010] and [100] crystallographic directions.

**Electron diffraction:**

Selected area electron diffraction (SAED) patterns were acquired using a Thermo Fisher Talos F200X microscope operated at 200 kV. Consistent SAED patterns were obtained from multiple regions.

**STEM imaging and post processing:**

The Cu-deficient sample exhibited beam sensitivity at 200 kV, leading to defect degradation; hence, atomic-resolution STEM and EDS analyses were conducted at a reduced accelerating voltage of 80 kV. Atomic-resolution ADF-STEM imaging was performed using a double spherical aberration–corrected Thermo Fisher Titan$^3$ G2 S/TEM. The Cu-deficient ADF-STEM image was acquired at a camera length of 115 mm with detector collection angles ranging from 42 to 244 mrad at 80 kV accelerating voltage.

These ADF-STEM images were drift-corrected by acquiring two images at 90 degrees with respect to each other. Drift correction was performed using MATLAB code developed by Ophus *et al.* [29], and a Gaussian blur with a 3.5-pixel radius was applied using in-house developed Easy STEM software in MATLAB [30].

The Cu-rich crystals are less beam sensitive compared to Cu-deficient crystals, so the STEM studies on Cu-rich crystals were performed at 300 kV. The ADF-STEM images of Cu-rich crystals were also acquired at a camera length of 115 mm with detector collection angles ranging from 42 to 244 mrad at 300 kV accelerating voltages. Similar post processing as Cu-deficient crystals including drift correction and a gaussian blur with a 2-pixel radius was applied on the Cu-rich ADF STEM images shown in Figure 2. Figure 4c and 4d are also drift corrected and with a gaussian blur with a 2.5-pixel radius applied on Figure 4d. The ADF-STEM images in Figure 3c and 3d are acquired at a camera length of 185mm with a detector collection angles ranging from 26 to 149 mrad.  In-built bandpass filtering in DigitalMicrograph software was used in SI Figure S5 to resolve atoms better along [100] zone axis.

Angle measurements between the needle shaped features in Figure S8 were performed using ImageJ, and the histogram was plotted using Python. The measured raw data were fitted using normfit() from scipy.stats to extract the mean and standard deviation. The error was calculated by dividing the standard deviation by $\sqrt{N}$, where N is the number of measurements. The range of angles mentioned= mean $\pm$ 2*standard deviation.

**EDS data processing:**

Image intensity sum per row analysis was performed on the EDS maps corresponding to Mn and Cu distributions in Cu-deficient sample and near stacking fault region in Cu-rich sample. Each image was normalized, and the total intensity was summed across all columns for every image row, generating one-dimensional profiles of intensity versus row index. The y-axis (row index) was inverted to correspond to the original image orientation.

EDS map analysis of needle shaped features was performed using in-house Python code. Intensity difference maps were generated from normalized elemental maps. Pearson correlation analysis was conducted on cropped images (x: 0–450 pixels, y: 50–450 pixels) to exclude edge regions, by calculating pixel-wise Pearson correlation coefficients between the cropped ADF-STEM image and each elemental map after flattening to 1D intensity vectors. The NumPy corrcoef function was used for correlation computation. Correlations were interpreted as $|r| < 0.1$ (no correlation), $r > 0.1$ (positive correlation), and $r < -0.1$ (inverse correlation).

**Unsupervised machine learning on sliding window FFT:**

ADF-STEM images were analyzed using an in-house developed Python code for local crystallographic segmentation. Sliding windows (128×128 pixels, 16×16 step) extracted FFT spectra, which were flattened, standardized (StandardScaler), and reduced via principal component analysis (PCA) and Gaussian mixture modeling (GMM; 3 components, full covariance) clustered the PC scores, yielding a cluster map overlaid on the original image.

**Strain mapping:**

Strain maps were obtained by applying geometric phase analysis to STEM images using the Strain++ software [31]. In this approach, two non-collinear reciprocal lattice vectors were selected, and the local phase variations of the corresponding Bragg reflections were converted into real space lattice displacement fields and, subsequently, into strain fields.

**EELS processing:**

EELS datasets were processed using the HyperSpy Python software package [32]. The core loss peak was aligned with respect to the zero loss peak, and the data were denoised by selecting the first five principal components in Hyperspy; the data were summed along each row to enhance the signal-to-noise ratio. The $Mn^{2+}$ reference was plotted from the literature data for MnO [33].

**Te interlayer distance measurement from STEM image:**

In-house Python codes were used to analyze the Te positions in the STEM images. The atomic positions were identified based on their maximum intensity, and the Te interlayer distances were measured from these Te positions.

**STEM image simulation:**

STEM image simulations were executed using the open-source abTEM package in Python [34]. The simulation parameters were chosen to closely match the experimental conditions. A $C_s$ of 0 nm, annular detector with a collection angle of 42 to 200 mrad, and 8 frozen phonon configurations with 0.1 Å standard deviation in atomic displacements to sample different thermal configurations were used to perform stem image simulations.

***DFT calculations:***

The crystal structures are given from the X-ray refinement. For partial occupancy, only the dominant element is assigned, except for the “additional sites” in the Cu-rich structure where Mn is assigned. The resulting structure has the formula $Mn_4Cu_4Si_4Te_{12}$ for Cu-deficient, and $Mn_4Cu_6Si_4Te_{12}$ for Cu-rich. Spin-polarized density functional theory calculations are done by VASP [35–38] with exchange-correlation treated at the PBE level[39] with rotationally invariant Hubbard U corrections [40] 3.9 eV for Mn and 7.2 eV for Cu d-orbitals, similar to our previous work[8]. The self-consistent field calculation is done with a cutoff energy of 500 eV, MP k-mesh [41] 7×4×7, convergence criterion $1\times10^{-7}$ eV, and Gaussian smearing with sigma 0.01 eV. The Born effective charge is calculated with the "LEPSILON" tag [42] in VASP for the X-ray-refined Cu-rich structure (containing no partial occupancy as discussed above), and the resulting charges are applied to all structures (containing partial occupancy) for the polarization estimation.

***XPS measurements:***

XPS measurements were performed at room temperature a Physical Electronics VersaProbe III instrument at Penn State. This system is equipped with a monochromatic Al Kα x-ray source (Energy= 1486.6 eV) with a concentric hemispherical analyzer. Here, the charge neutralization was performed using both low energy electrons (<5 eV) and argon ions. Peaks were charge referenced to $CH_x$ band in the carbon 1 s spectra at 284.8 eV. Measurements were made at a takeoff angle of 45° with respect to the sample surface plane. High-energy resolution spectra were collected using a pass energy of 69.0 eV with a step size of 0.25 eV.

**Acknowledgements**

N.A., W.X., S.A., and S.V.G.A. acknowledge support from the Experimental Condensed Matter Physics program, Division of Materials Sciences and Engineering, Basic Energy Sciences, U.S. Department of Energy (DOE), grant DE-SC0024943, for structural studies using STEM and XRD on Cu-rich samples. S.V.G.A. and N.A. also acknowledge partial support from the National

Science Foundation (NSF) through the Pennsylvania State University Materials Research Science and Engineering Center (MRSEC), award DMR-2011839 (2020–2026), for the structural study of Cu-deficient samples. B.Z. and V.H.C. acknowledge support from the Two-Dimensional Crystal Consortium–Materials Innovation Platform (2DCC-MIP) under NSF cooperative agreement DMR-2039351.

**Supplemental Information**

# Impact of Cu–Mn ratio on Structure and Defects in Layered Multiferroic $Cu_{1-x}Mn_{1+y}SiTe_3$

Sai Venkata Gayathri Ayyagari[1], Boyang Zheng[2,3], Sreekant Anil[1], Subrata Ghosh[2,3], Yuxi Zhang[1], Yu Liu[2,3], Chandan De[2,3], Ke Wang[4], Jeffrey Shallenberger[4], Weiwei Xie[5], Vincent H. Crespi[2,3], Zhiqiang Mao[1,2,3], and Nasim Alem[1*]

[1]Department of Materials Science and Engineering, The Pennsylvania State University, University Park, PA 16802, USA

[2]2D Crystal Consortium, Materials Research Institute, The Pennsylvania State University, University Park, PA 16802, USA

[3]Department of Physics, The Pennsylvania State University, University Park, PA 16802, USA

[4]Materials Research Institute, University Park, PA 16802, USA

[5]Department of Chemistry, Michigan State University, East Lansing, MI 48864, USA

* Corresponding author: nua10@psu.edu

**Figure S1-S3: Structural refinement using single crystal X-ray**

The single-crystal X-ray diffraction reciprocal-space reconstructions reveal composition-dependent structural disorder across the series. The Cu-deficient sample exhibits predominantly sharp and well-defined Bragg reflections, indicating good long-range periodic order, with anisotropic diffuse streaking observed in selected reciprocal-space sections. The directional character of the diffuse intensity suggests stacking disorder confined to one crystallographic axis (Figure S2), likely arising from imperfect interlayer registry within the layered framework, consistent with TEM observations shown in Section 2.1. The Cu-rich sample displays well-defined Bragg reflections characteristic of single-crystal behavior (Figure S3), accompanied by weak powder arcs and localized diffuse features in selected projections, suggesting moderate stacking disorder and minor secondary crystallites.

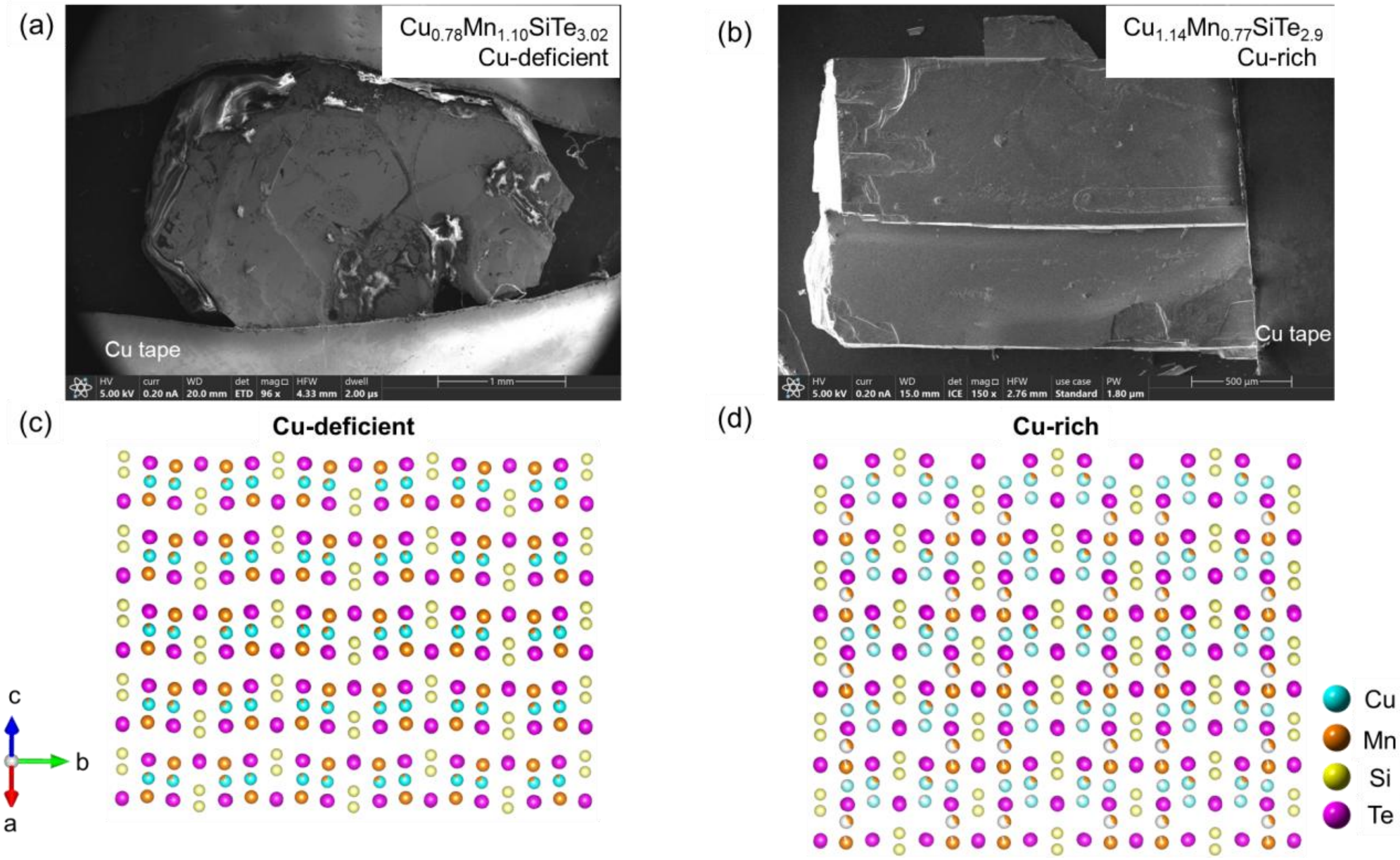

**Figure S1. Overview of crystals and structure refinement using single crystal X-ray:** (a, b) SEM images of Cu-deficient ($Cu_{0.78}Mn_{1.10}SiTe_{3.02}$) and Cu-rich ($Cu_{1.14}Mn_{0.77}SiTe_{2.9}$) crystals,

respectively. (c,d) X-ray-refined crystal structures of Cu-deficient and Cu-rich crystals, respectively.

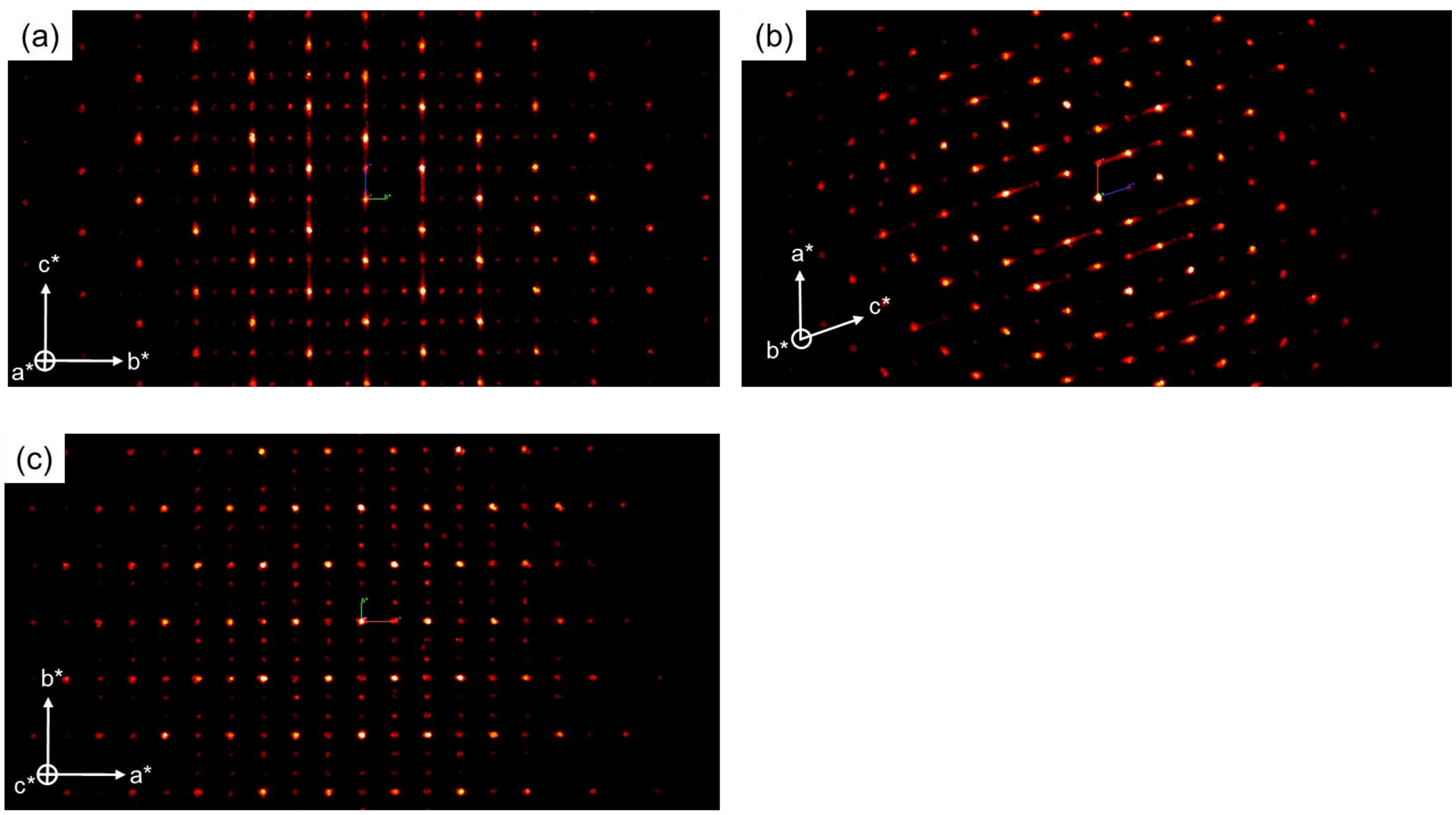

**Figure S2. Reciprocal-space reconstruction of the Cu-deficient sample ($Cu_{0.79}Mn_{1.08}SiTe_{2.97}$):** (a-c) Reconstructions along the [0kl], [h0l], and [hk0] directions, respectively. Sharp Bragg reflections confirm long-range periodic order, while weak directional diffuse streaking indicates anisotropic stacking disorder.

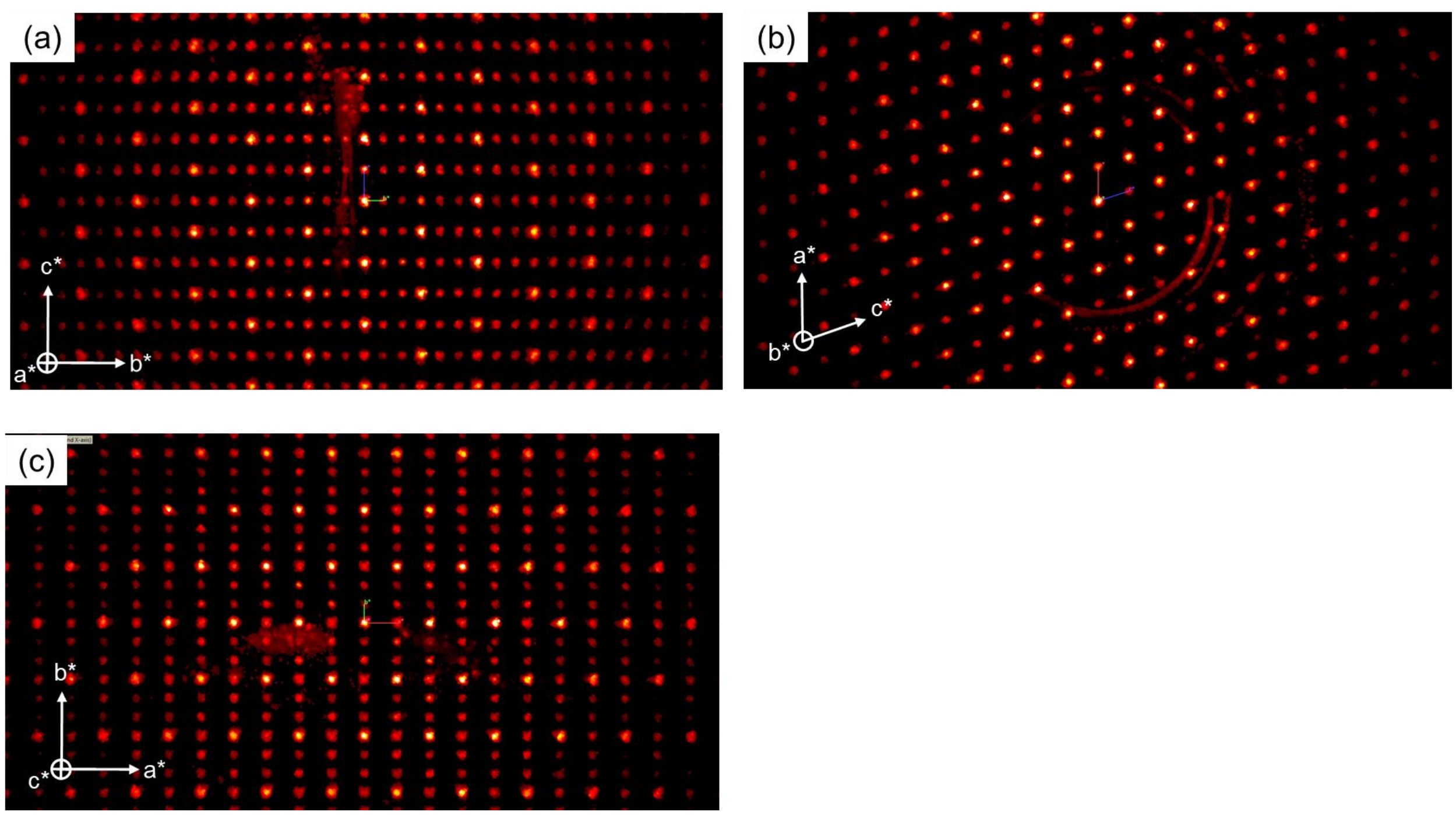

**Figure S3. Reciprocal-space reconstruction of the Cu-rich sample ($Cu_{1.14}Mn_{0.77}SiTe_{2.90}$):** (a-c) Reconstructions along the [0kl], [h0l], and [hk0] directions, respectively. Well-defined Bragg reflections confirm single-crystal character with moderate diffuse features.

**Figure S4: Understanding the structure of stacking fault like defect in Cu-deficient crystals using unsupervised machine learning algorithms**

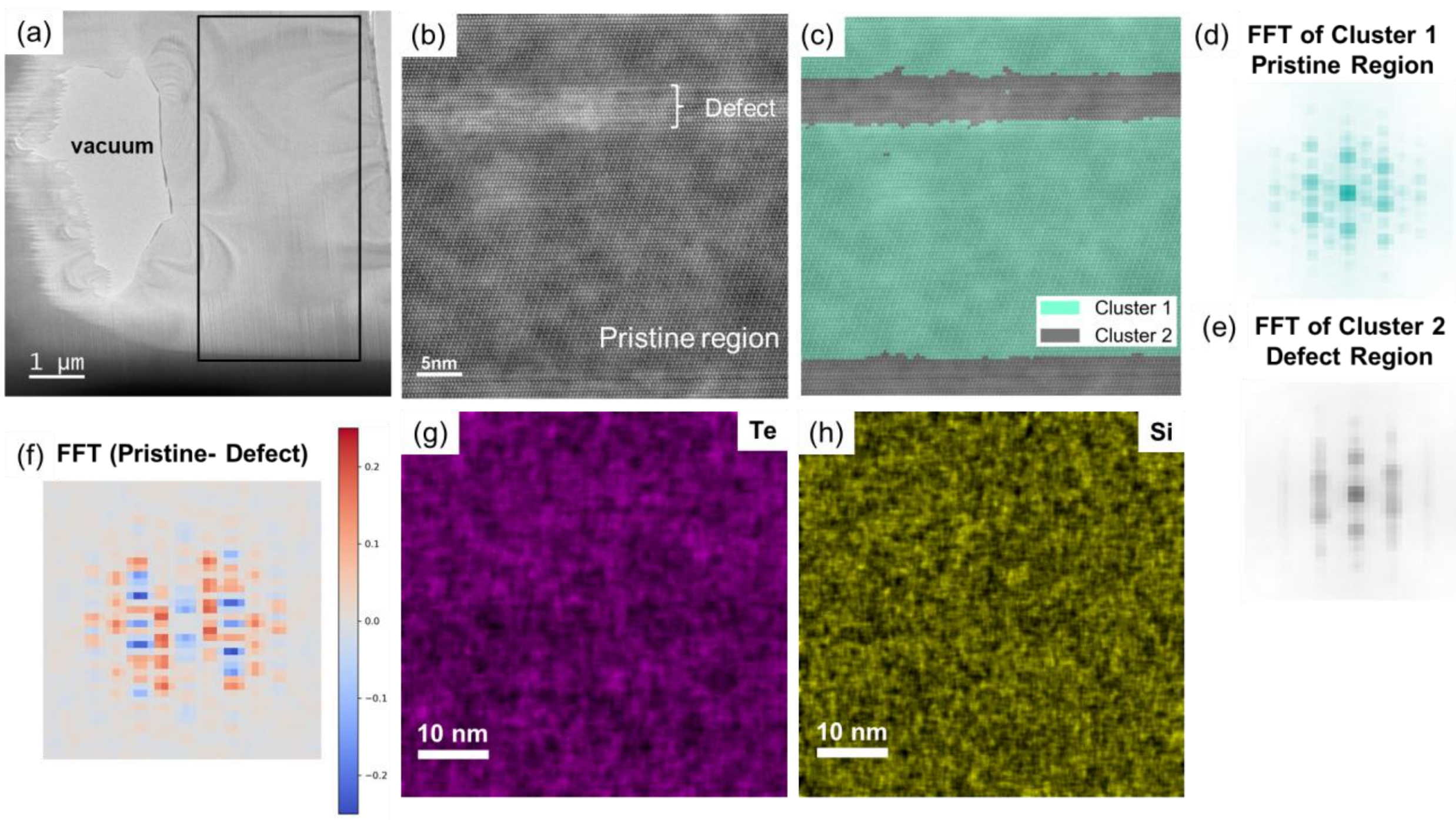

**Figure S4. Analysis of stacking fault defect regions in Cu-deficient sample:** (a) Low magnification TEM image with rectangle highlighting the numerous stacking faults present in the Cu-deficient crystal. (b) ADF-STEM image and the corresponding cluster map (c) obtained after performing principal component analysis (PCA) on sliding FFTs generated using a 128 x 128 pixel window, moved with a step size of 16 x 16 pixels along both x and y direction. A Gaussian mixture model was applied to the FFTs assuming two distinct clusters. (d-f) Log-averaged FFTs from each cluster region highlighting the differences between the defect region and pristine region. The difference between the FFTs of cluster 1 (Cu-rich defect) and cluster 2 (pristine), shown in (f), indicates a change in symmetry. (g-h) EDS maps of Te and Si, respectively.

**Figure S5: Oxidation state analysis of Mn and structure along [100] zone**

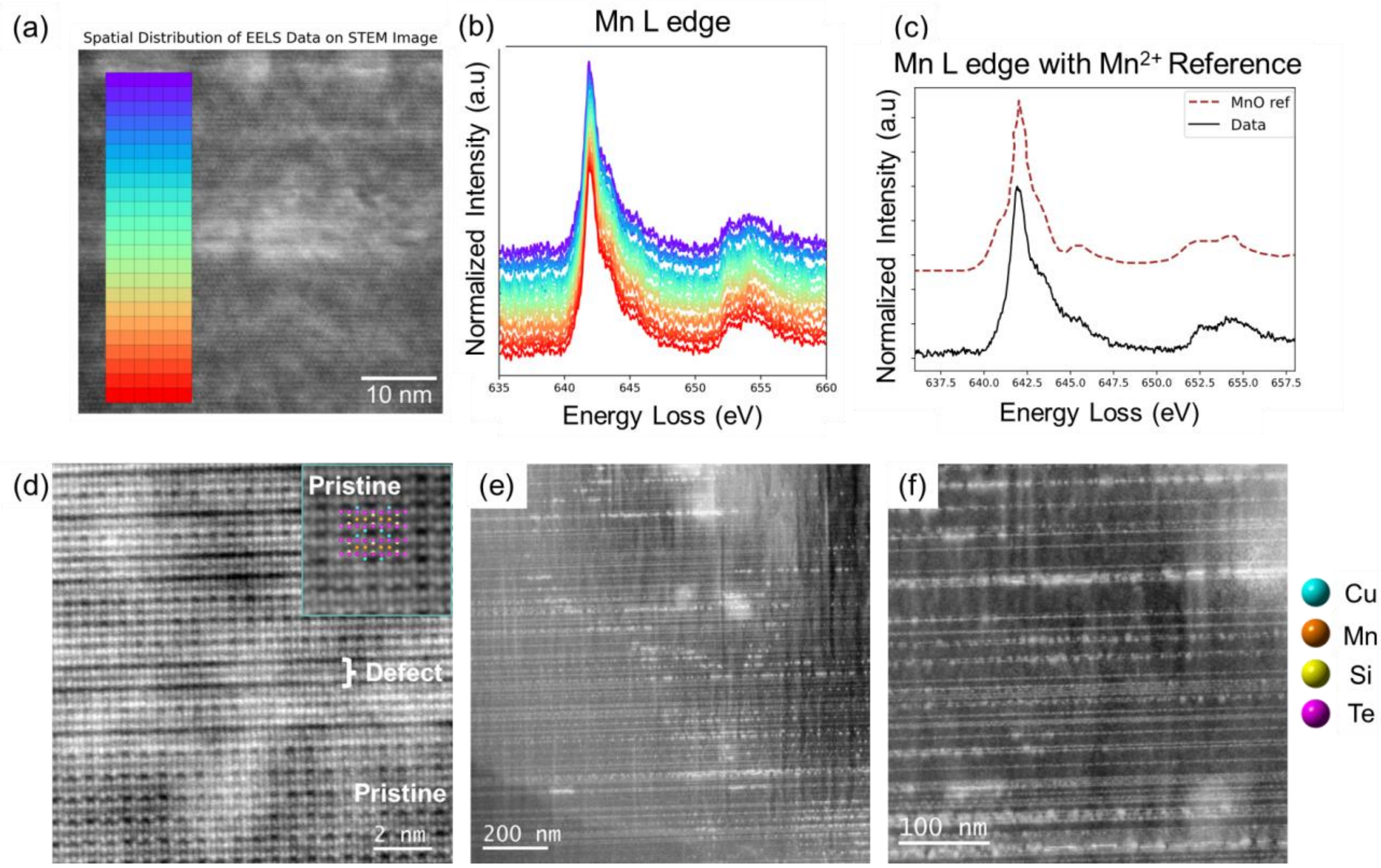

**Figure S5. Oxidation state and structure investigation in Cu-deficient sample:** (a-b) STEM image overlaid with the probe positions corresponding to the Mn L edge EELS data shown in (b). (c) Summed up EELS data of Mn L edge along with the reference $Mn^{2+}$ L edge from literature[1]. (d) STEM image (bandpass filtered) of pristine and defect regions acquired along [100] zone axis with inset showing a magnified region of pristine area overlaid with the structure model. (e-f) Beam induced damage on the Cu-deficient sample along [100] zone axis. It is evident that the stacking faults are being damaged first as the electron beam interacts with the sample. We also observed that sample along [100] damaged in electron beam at a faster rate than [010] zone sample at 80 kV accelerating voltage.

**Figure S6: Strain Maps of these defect regions in Cu-deficient sample**

Geometric Phase Analysis (GPA) was implemented to calculate strain maps across the defect regions in Cu-deficient sample.

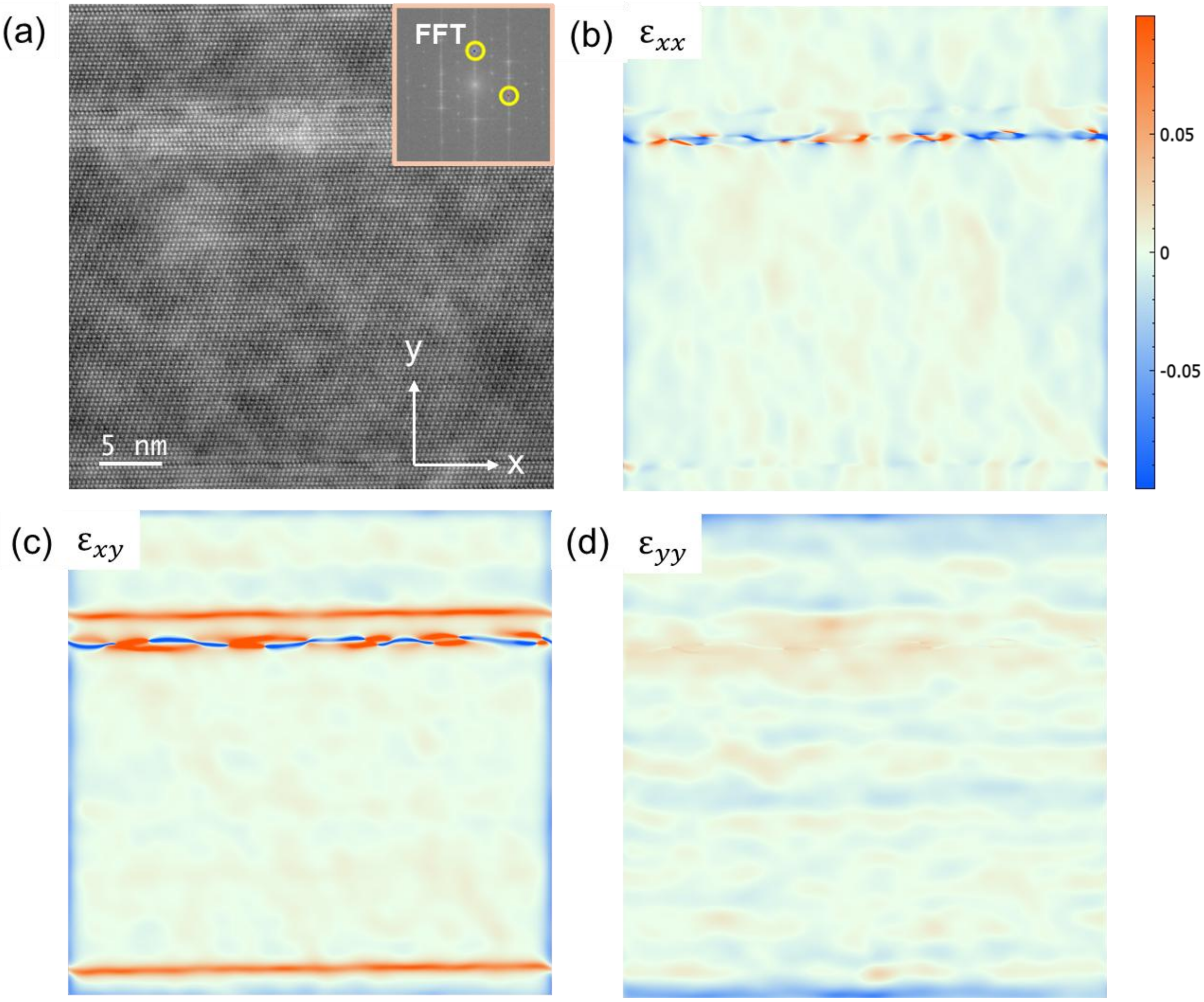

**Figure S6. Strain mapping of stacking faults in Cu-deficient sample:** (a) ADF-STEM image of $Cu_{0.78}Mn_{1.10}SiTe_{3.02}$ cross-section specimen acquired along the [010] zone axis, with an inset FFT showing the two non-collinear reflections selected for GPA analysis. (b-d) Corresponding strain maps $\varepsilon_{xx}$, $\varepsilon_{xy}$, $\varepsilon_{yy}$, respectively.

**Figure S7: Models of Te layer stacking order of defect regions in Cu-deficient sample**

Using density functional theory (DFT), we investigated the structure of the stacking faults, that is, the AB-stacking of the Te-layers, while the pristine region exhibits ABC-stacking (Figure S7). Considering that the Cu-deficient sample contains more Mn than Cu, and the stacking fault region is relatively Cu-rich and Mn-deficient, we hypothesize that this local compositional variation drives the formation of stacking faults.

To explore this mechanism, structural models for both regions are required. For the pristine region of the sample, which exhibits ABC-stacking, we use the structure obtained from single crystal X-ray refinement as previously reported [2]. In the ideal composition $Cu_4Mn_4Si_4Te_{12}$ composition (Cu:Mn ratio = 1), the ABC-stacking is 0.99 eV/u.c. more stable than the AB-stacked structure. Here the AB-stacking structure is created by modifying the lattice and Cu positions while maintaining the cell volume and the Si-dimer configuration (Figure S7).

To computationally take into account the fact that Cu:Mn ratio $<1$, a Cu-deficient structure is needed and can be modeled through replacing two Cu atoms by one Mn atom, yielding the composition $Cu_2Mn_5Si_4Te_{12}$ (Cu:Mn ratio $< 1$), which preserves the valence electron counting and the semiconducting nature of the material. The AB-stacked structure of this composition is constructed in the same manner as described above. In this case, the ABC-stacking is 0.13 eV/u.c. more stable than the AB-stacking. These results indicate that ABC stacking is more stable than AB stacking. Furthermore, these findings align with our STEM observations, which show that the stacking-fault regions (AB stacking) tend to degrade first under electron-beam irradiation compared to the pristine region (ABC stacking).

For the stacking fault region, which is comparatively Cu-rich, the atomic structure remains unclear, as the STEM images do not provide a resolvable distinction between the Mn and Cu arrangements within the Te layers. Simply adding one Cu atom to the metal layer of the unit cell, resulting in $Cu_5Mn_4Si_4Te_{12}$ composition, yields a structure in which ABC-stacking 0.83 eV/u.c. more stable than the AB-stacking. This indicates that such a composition would still energetically favor the ABC stacking, similar to the pristine region, and therefore cannot account for the stacking-fault structure based purely on energetic considerations. Further calculations, potentially including the kinetic effect, combined with further characterization in this region, are necessary to fully understand the atomic configuration of stacking faults and their impact on properties.

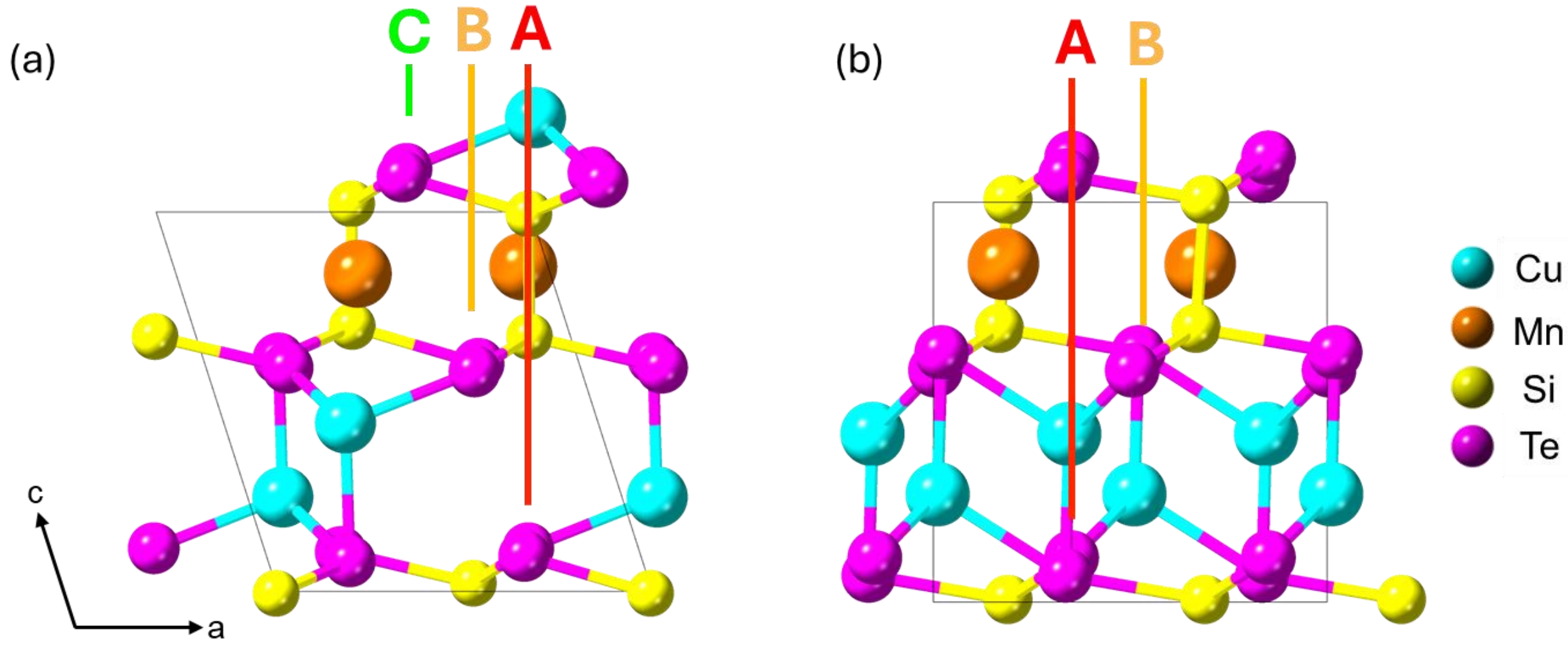

**Figure S7. Models of Te-layer stacking orders**. **(a)** The model for the major structure of the Cu-deficient sample with an ideal (Cu : Mn ratio = 1) $CuMnSiTe_3$ composition, which shows the ABC-stacking of Te-layers. **(b)** One of the candidate structures being tested for the stacking fault regions, which shows the AB-stacking of Te-layers.

**Figure S8: Angle measurement between the needle shaped features in Cu-rich ($Cu_{1.14}Mn_{0.77}SiTe_{2.9}$) crystals**

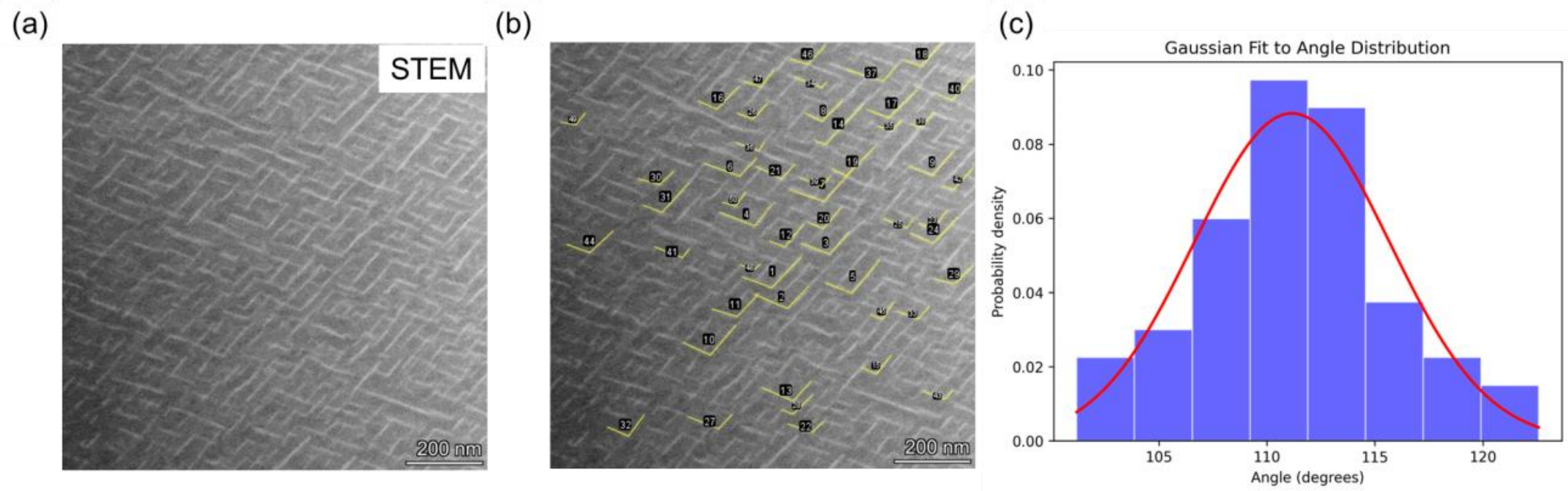

**Figure S8: Angle measurement between the needle shaped features in Cu-rich sample:** (a) STEM image. (b) Angle measurements performed using ImageJ on STEM image (a). (c) Histogram of measured angles (mean= 111°, standard deviation= 4.5° with error of 0.6°).

**Figure S9: Needle-shaped features in Cu-rich crystals observed along [100] zone axis**

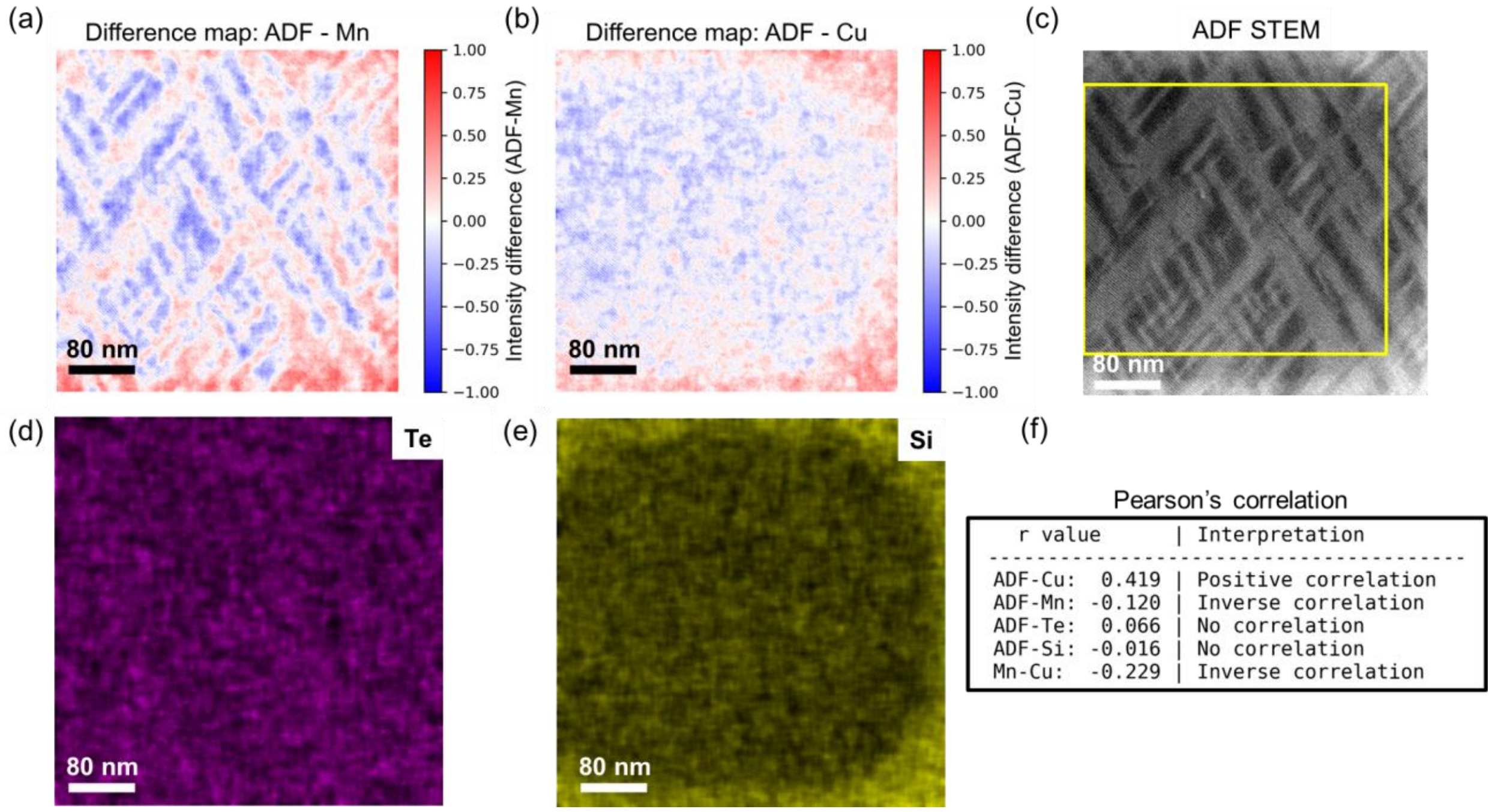

**Figure S9. Elemental distribution maps analysis of needle-shaped features in Cu-rich crystals:** (a-b) Intensity difference map between the ADF-STEM image and Mn and Cu maps, respectively, highlighting that the bright regions in STEM image represent Mn-deficient and Cu-rich regions. (c) Selected area of the ADF-STEM image (yellow rectangle) used to perform Pearson correlation between the ADF-STEM image and elemental maps shown in (f). Regions that are Mn-deficient are Cu-rich and appear as bright regions in the ADF-STEM image. (d-e) EDS maps of Te and Si, respectively.

**Figure S10: Understanding the structure of needle-shaped features in Cu-rich crystals using unsupervised machine learning algorithms**

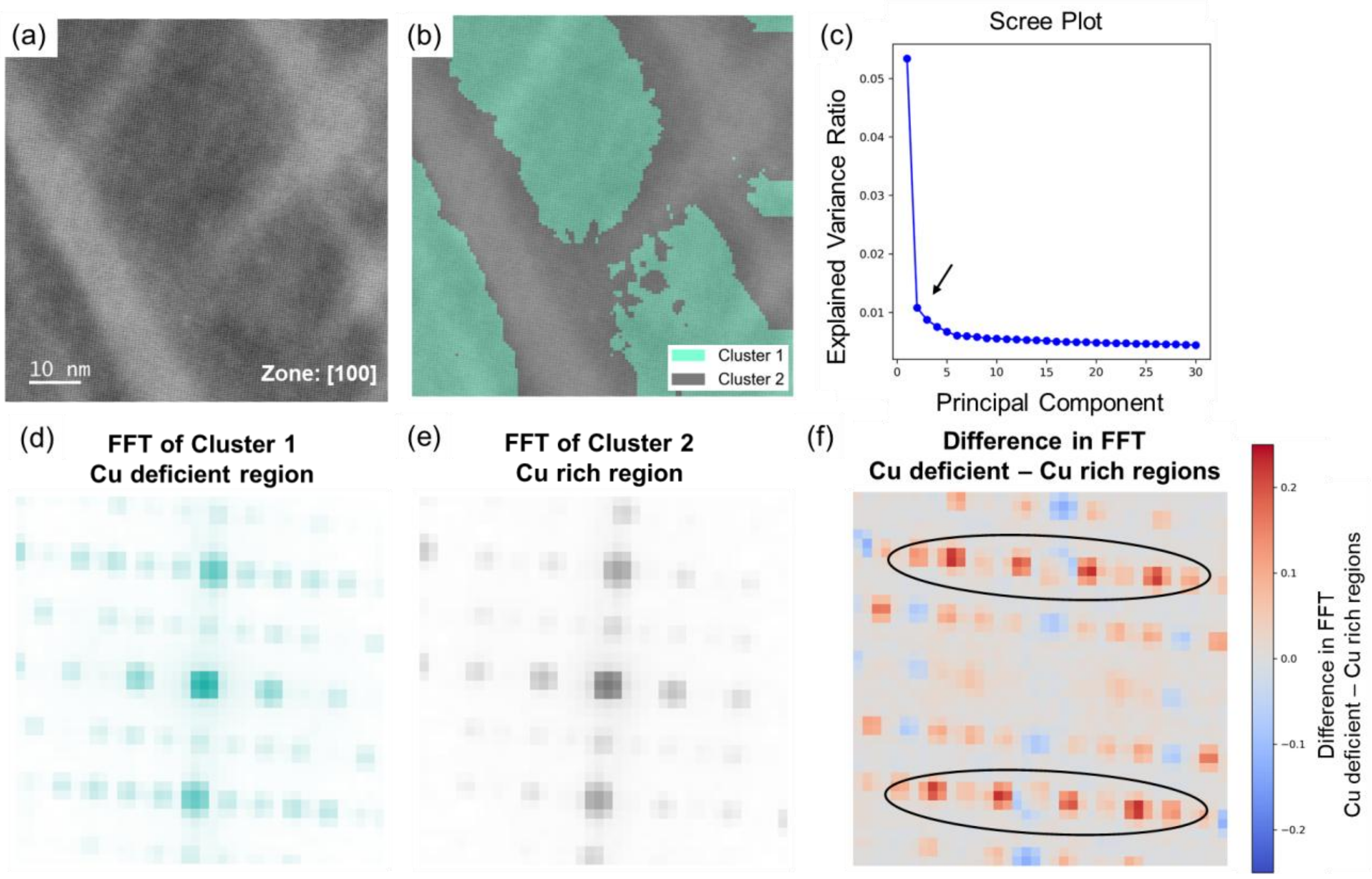

**Figure S10. Analysis of needle shaped features in Fourier space using unsupervised machine learning algorithms on sliding window FFT:** (a) ADF-STEM image shown in Figure 3c and the corresponding cluster map (b) obtained after performing principal component analysis (PCA) on sliding FFTs generated using a 128 x 128 pixel window, moved with a step size of 16 x 16 pixels along both x and y direction. A Gaussian mixture model was applied to the FFTs assuming two distinct clusters determined from PCA Scree Plot shown in (c). (d-e) Log-averaged FFTs from each cluster region highlighting the differences between the needle shaped defect region (Cu rich region) (grey color) and matrix region (Cu deficient region). The difference between the FFTs, shown in (f), indicates a change in symmetry, with the inset highlighting the relative differences in the periodicity of reflections.

**Figure S11: STEM images on the needle shaped defect regions in Cu-rich samples**

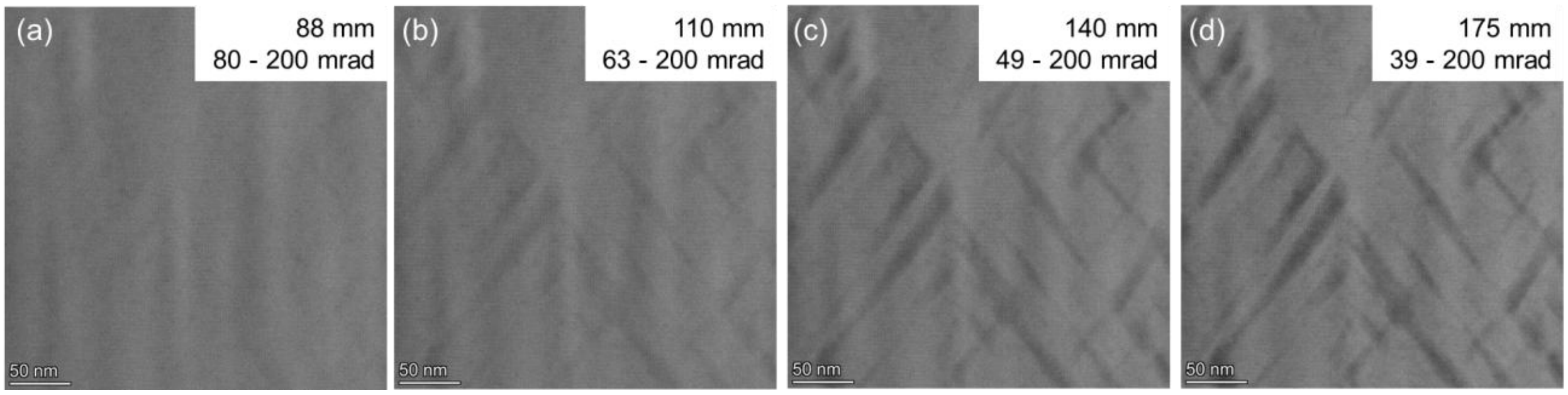

**Figure S11. STEM images of the Cu-rich sample at different camera lengths (collection angles) exhibit diffraction contrast from these features:** (a-d) STEM images acquired at camera lengths ranging from 88 mm to 175 mm show a transition from Z-contrast imaging to diffraction contrast imaging.

**Figure S12: Strain Maps of these needle shaped defect regions in Cu-rich samples**

Geometric Phase Analysis (GPA) was implemented to calculate the strain maps across the defect regions in Cu-rich samples.

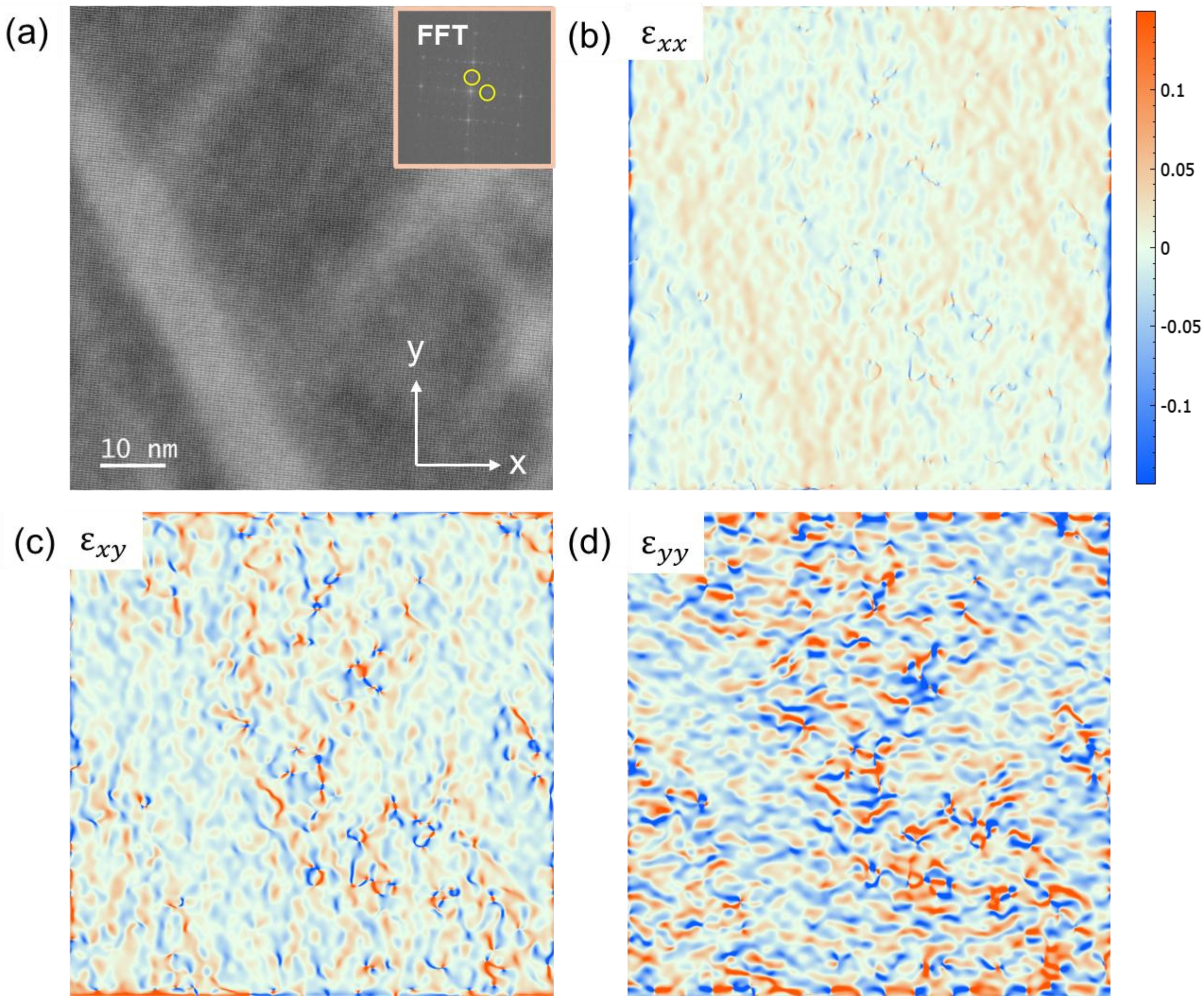

**Figure S12. Strain mapping of needle-shaped features in Cu-rich sample:** (a) ADF-STEM image of $Cu_{1.14}Mn_{0.77}SiTe_{2.9}$ cross-section specimen acquired along the [100] zone axis, with an inset FFT showing the two non-collinear reflections selected for GPA analysis. (b-d) Corresponding strain maps $\varepsilon_{xx}, \varepsilon_{xy}, \ \varepsilon_{yy}$, respectively.

**Figure S13: Alternative candidate structures for Cu-rich samples from density functional theory**

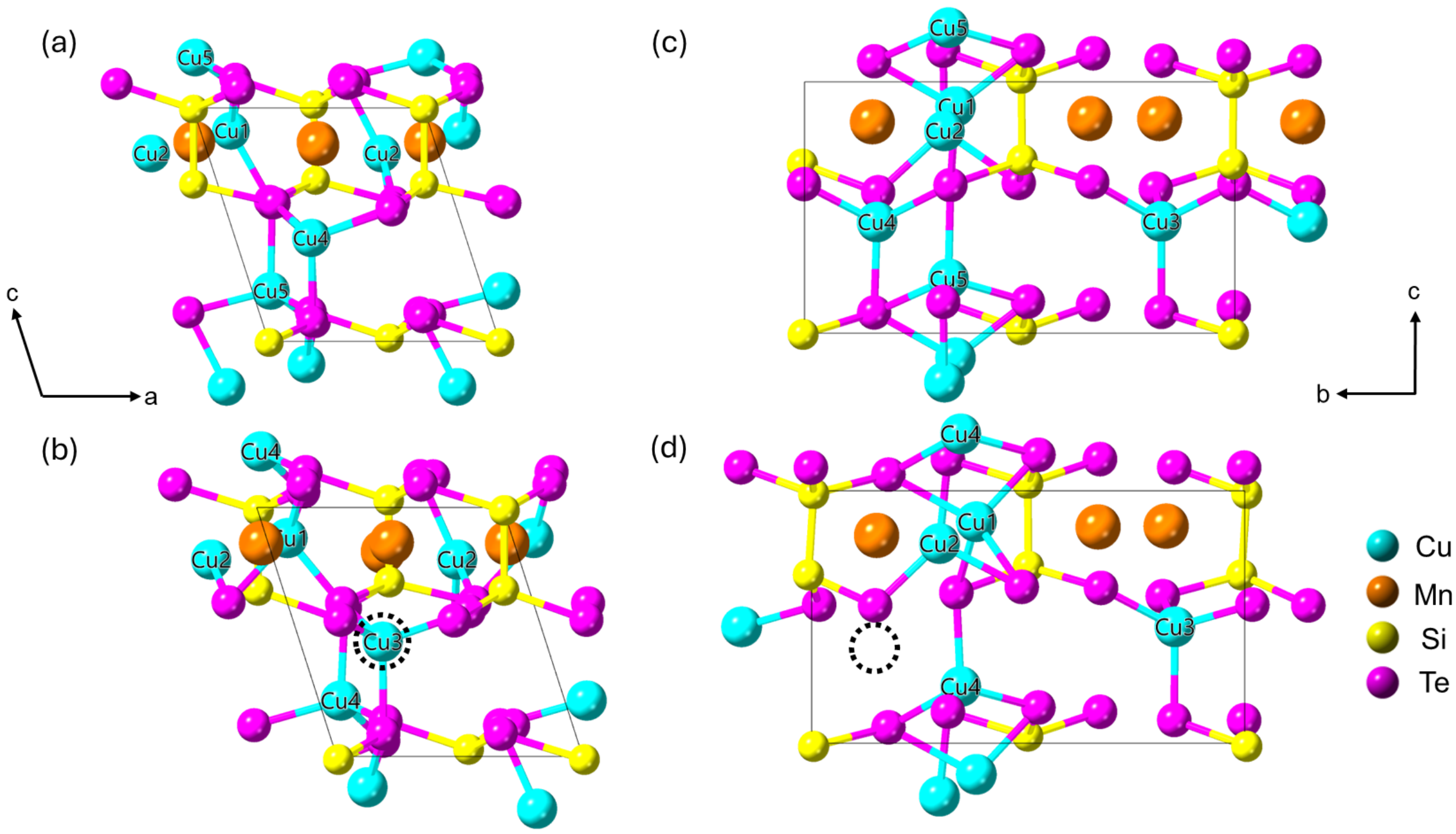

**Figure S13. Side views for alternative candidate structures for the Cu-rich samples from DFT:** (a) and (c) are the [010] and [100] views for one structure ($Mn_3Cu_5Si_3Te_{12}$), respectively. (b) and (d) are for another structure ($Mn_3Cu_4Si_3Te_{12}$) which has one fewer Cu (dashed circle in (b, d)) in the unit cell. The missing Cu atom is not evident along [010] zone axis (Figure S13b) as eclipsed by Cu3. This is consistent with our experimental results, as we observe these needle-shaped features only along [100] zone axis and not [010] zone axis. Both structures are relaxed under DFT with fixed lattice vectors from the X-ray refinement. A 1:1 ratio average of these two structures would give $Cu_{1.13}Mn_{0.75}SiTe_3$. It should be noted that both structures should be disordered in the real sample, meaning that the replacement of Mn by the Cu-dimer could happen to any of the Mn in the unit cells. The ratio for Te-layer distances (Si-dimer : Cu-layer) is 1.080 for (a) and 1.091 for (b). The biggest difference between these two structures proposed by DFT from the X-ray-

refined structure is that the Cu atoms in the Si-dimer layer are not at the centres of octahedra; consequently, much shorter Cu-Cu distances are found in the Si-dimer layer here. Specifically, for the (a) structure, we have distances Cu(1)-Cu(2) = 2.57 Å, Cu(1)-Cu(5) = 2.51 Å, and for the (b) structure, Cu(1)-Cu(2) = 2.51 Å, Cu(1)-Cu(4) = 2.65 Å, indicating strong Cu-Cu interaction.

**Figure S14: Te-layer distance measurement for Cu rich samples from STEM image**

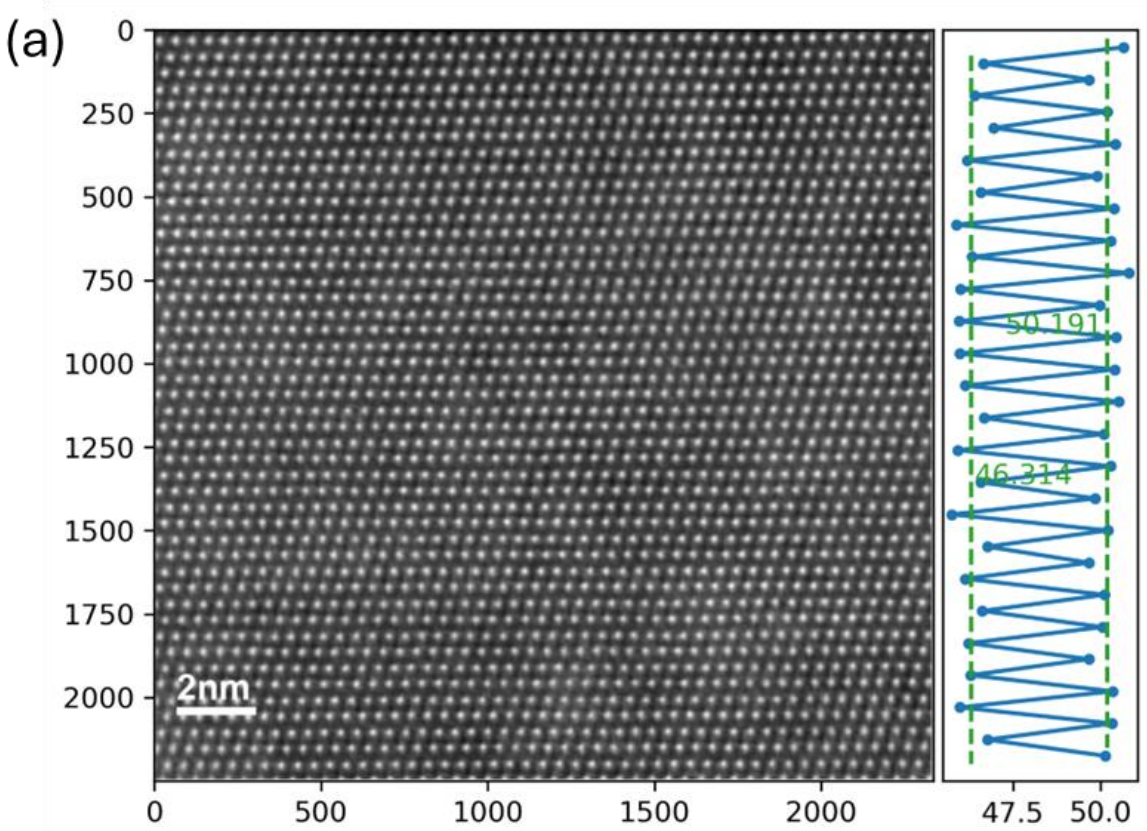

**Figure S14. Te-layer distances analysis for Cu-rich sample along [010] zone axis**: (a) Te layer distance measured in the STEM image shown in Figure 2f. The distances are measured in pixels. The structure shows alternating Te-layer distances (green-labelled region ~50.19 and ~46.31 pixels), with one about 8% larger than the other, meaning a ratio between the layer distances of ~1.08.

**Figure S15: Understanding the structure of stacking fault like defect in Cu-rich crystals using unsupervised machine learning algorithms**

Unsupervised machine learning applied to the sliding window FFTs across the STEM image (Figure S15) identifies pristine regions matching the overall structure identified from single crystal XRD alongside defect regions showing Te stacking changes, with sections showing Cu-rich and Mn rich segregation, and local symmetry changes.

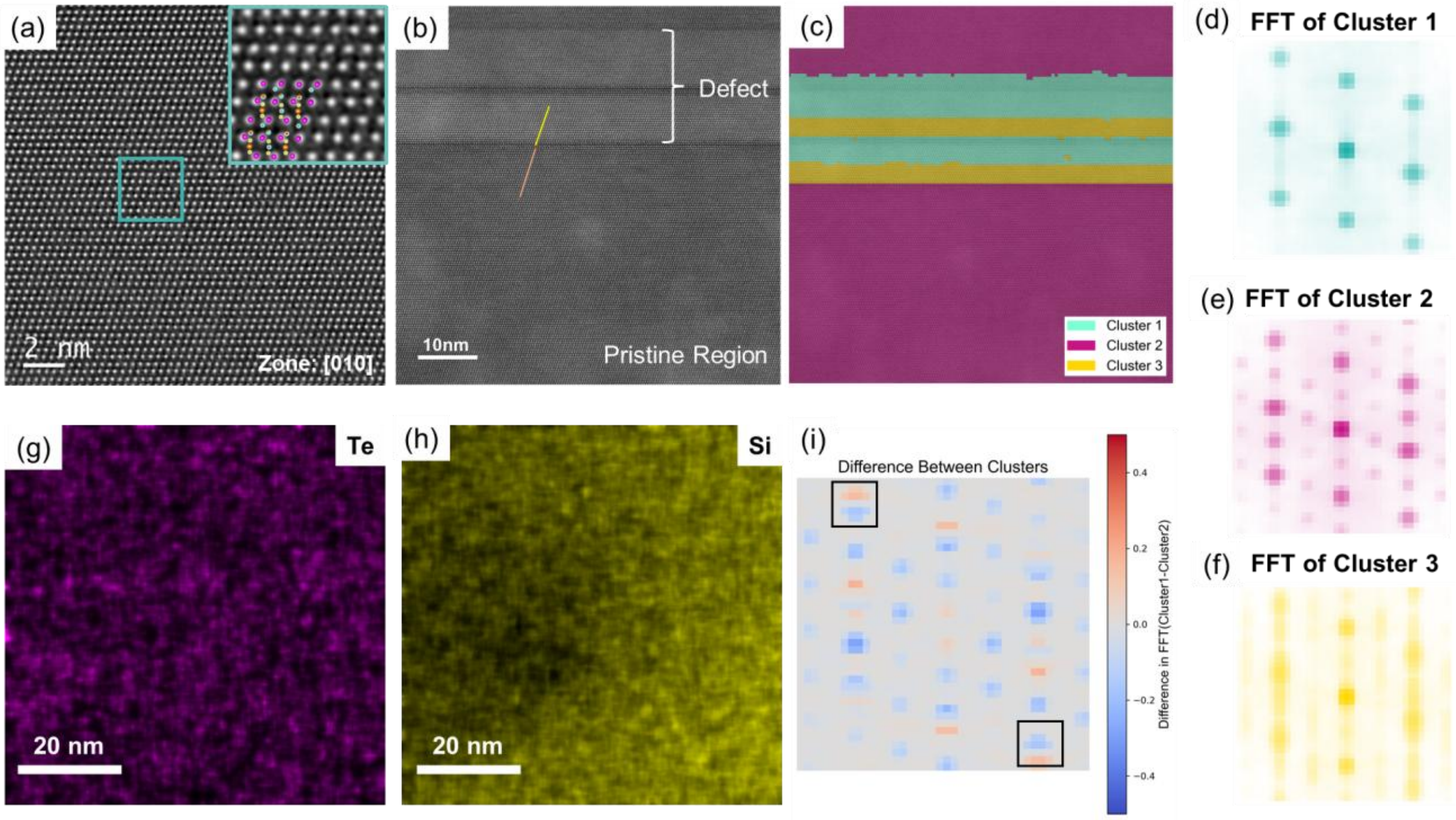

**Figure S15. Analysis of stacking fault like defect regions:** (a) Bandpass filtered ADF-STEM image of the pristine region along [010] zone axis showing the atomic arrangement consistent with the structure model obtained from single crystal XRD. (b) ADF-STEM image shown in Figure 4c and the corresponding cluster map (c) obtained after performing principal component analysis (PCA) on sliding FFTs generated using a 128 x 128 pixel window, moved with a step size of 16 x 16 pixels along both x and y direction. A Gaussian mixture model was applied to the FFTs

assuming three distinct clusters. The resulting cluster map does not accurately capture the Cu-rich and Mn-rich regions within the defect area due to the chosen window and step sizes. (d-f) Log-averaged FFTs from each cluster region highlighting the differences between the defect region and pristine region. The difference between the FFTs of cluster 1 (Cu-rich defect) and cluster 2 (pristine), shown in (i), indicates a change in symmetry, with the black box inset highlighting relative changes in the position of the similar reflection. (g-h) EDS maps of Te and Si, respectively.

**Figure S16: Strain Maps of these defect regions in Cu-rich samples**

Geometric Phase Analysis (GPA) was implemented to calculate the strain maps across the defect regions in Cu-rich samples.

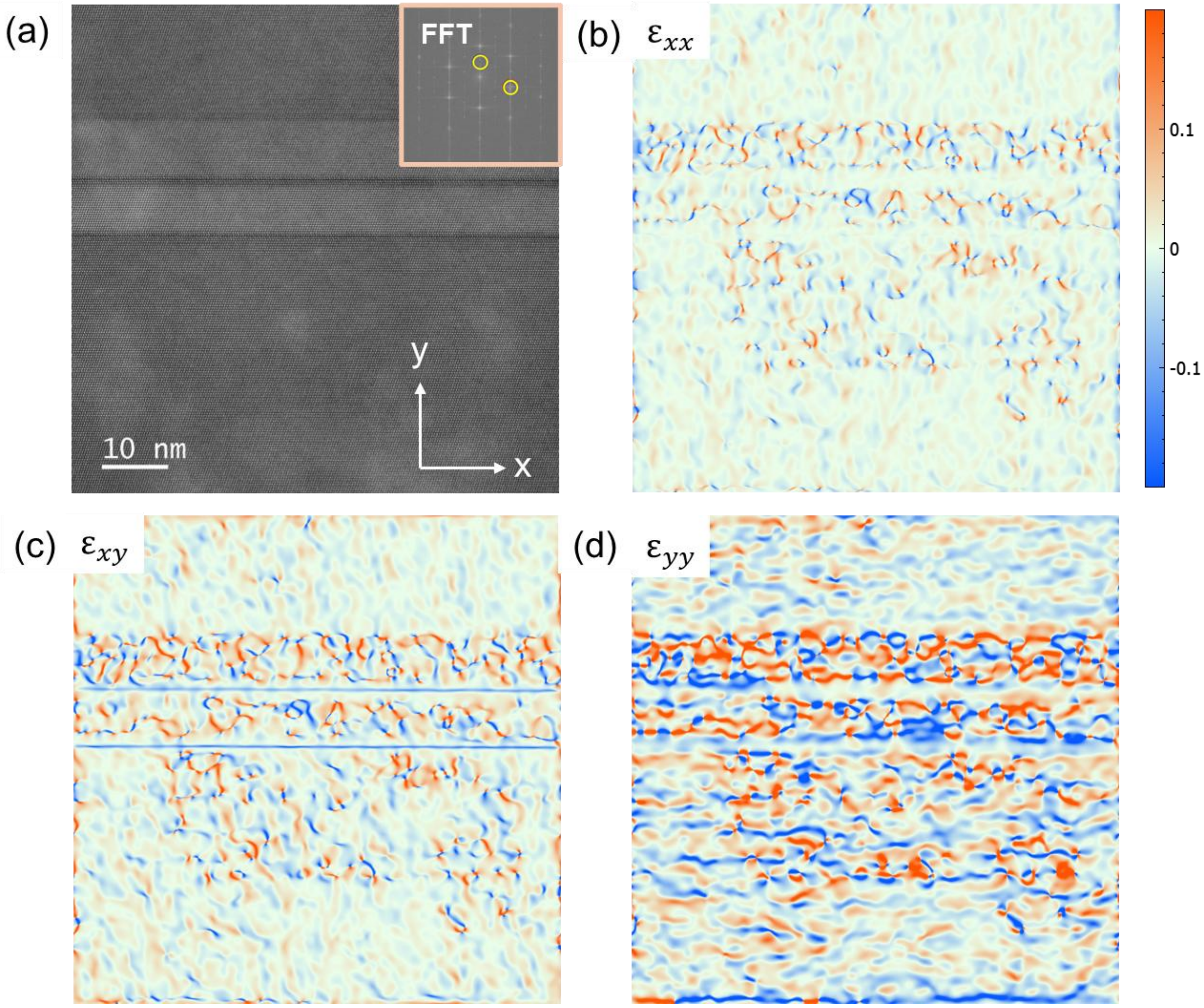

**Figure S16. Strain mapping of stacking fault type defect in Cu-rich sample:** (a) ADF-STEM image of $Cu_{1.14}Mn_{0.77}SiTe_{2.9}$ cross-section specimen acquired along the [010] zone axis, with an inset FFT showing the two non-collinear reflections selected for GPA analysis. (b-d) Corresponding strain maps $\varepsilon_{xx}, \varepsilon_{xy}, \varepsilon_{yy}$, respectively.

**Figure S17: Understanding the structure of stacking fault type defect region in Cu-rich crystals**

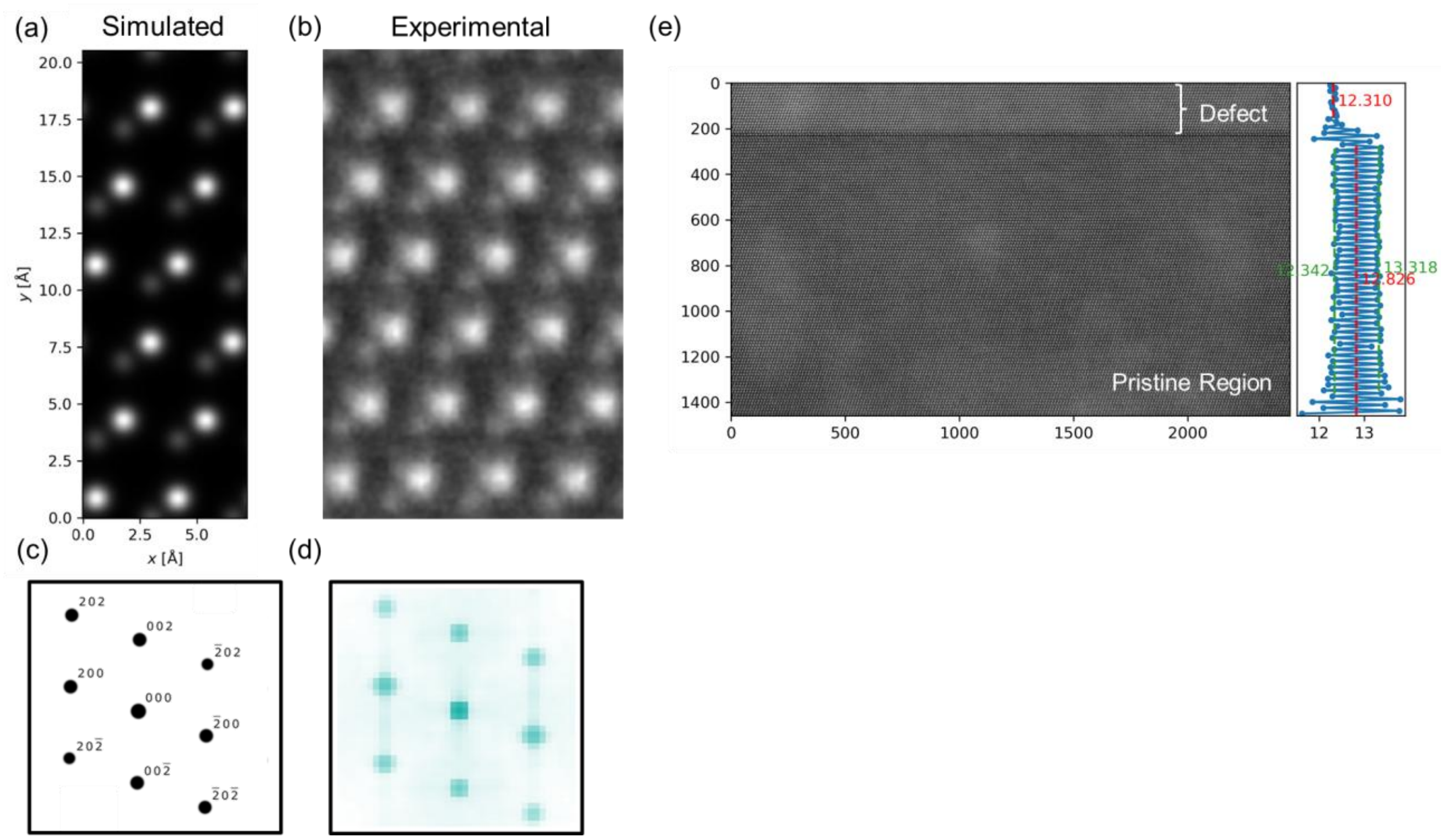

**Figure S17. Stacking-fault type defect in Cu-rich crystal:** (a,c) Simulated ADF STEM and electron diffraction pattern of proposed defect region's structure along [010] zone ($Cu_2SiTe_3$, mp-675120). (b,d) Experimental ADF STEM image (zoomed in from Figure 4d) and corresponding FFT (from Figure S15d). (e) Te interlayer distance measurement (similar to Figure S14) of a section of Figure 4c, highlighting that the defect region in loses the alternating Te atomic arrangement observed in pristine region and corresponds to an average Te-layer distance of 3.44 Å, matching with the proposed structure.

**Figure S18: Stacking fault in cleaved crystal sample of Cu rich crystal**

To determine if stacking fault like defects were restricted to the top regions of the crystal, a cleaved crystal was examined as shown in Figure S18. TEM studies on the cleaved crystal revealed that stacking faults are present in the middle of the sample and not exclusively near the top, indicating that neither FIB processing nor surface interactions are responsible for triggering stacking fault formation. Instead, the stacking faults are intrinsic to the crystal, with the chemical distribution of both Mn and Cu remaining the main driving forces behind their formation.

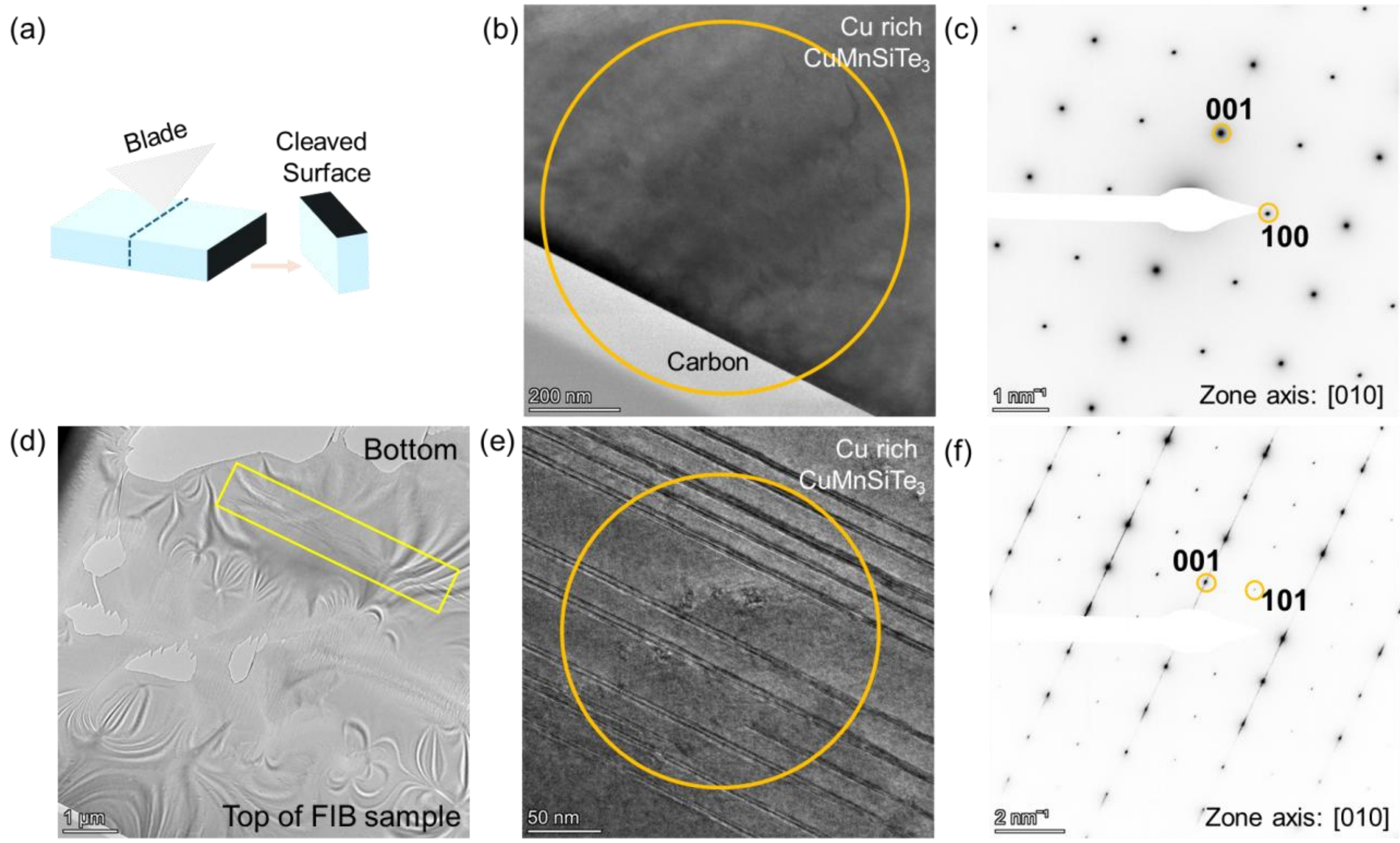

**Figure S18. Cu-rich cleaved crystal investigation:** (a) Schematic of a cleaved sample prepared using a blade, with a cross-sectional FIB lamella made from the cleaved crystal. (b) Selected area region from the top part of the FIB sample (region close to the carbon deposit), with a circular inset indicating the selected area corresponding to the electron diffraction pattern shown in (c). (d)

Lower magnification image of the FIB sample revealing a stacking fault in the middle of the lamella, highlighted by the yellow box. (e) Selected area region from the stacking-fault-like region in the sample, with a circular inset indicating the selected area corresponding to the electron diffraction pattern shown in (f).

**Figure S19 and S20: Loop-like features in Cu-rich crystals observed along both [010] and [100] zone samples**

In Cu-rich crystals, additional loop-like features are observed. DF-TEM images show that the contrast of these features depends on the reflection used to generate the image, suggesting they may be ferroelectric loops. STEM images acquired at different collection angles also exhibit strong diffraction contrast from these features. We hypothesize that this material could have closed or tubular domains in 3D which are projected as loops in the TEM, but this needs further investigation. EDS maps confirm there is no chemical inhomogeneity in these regions.

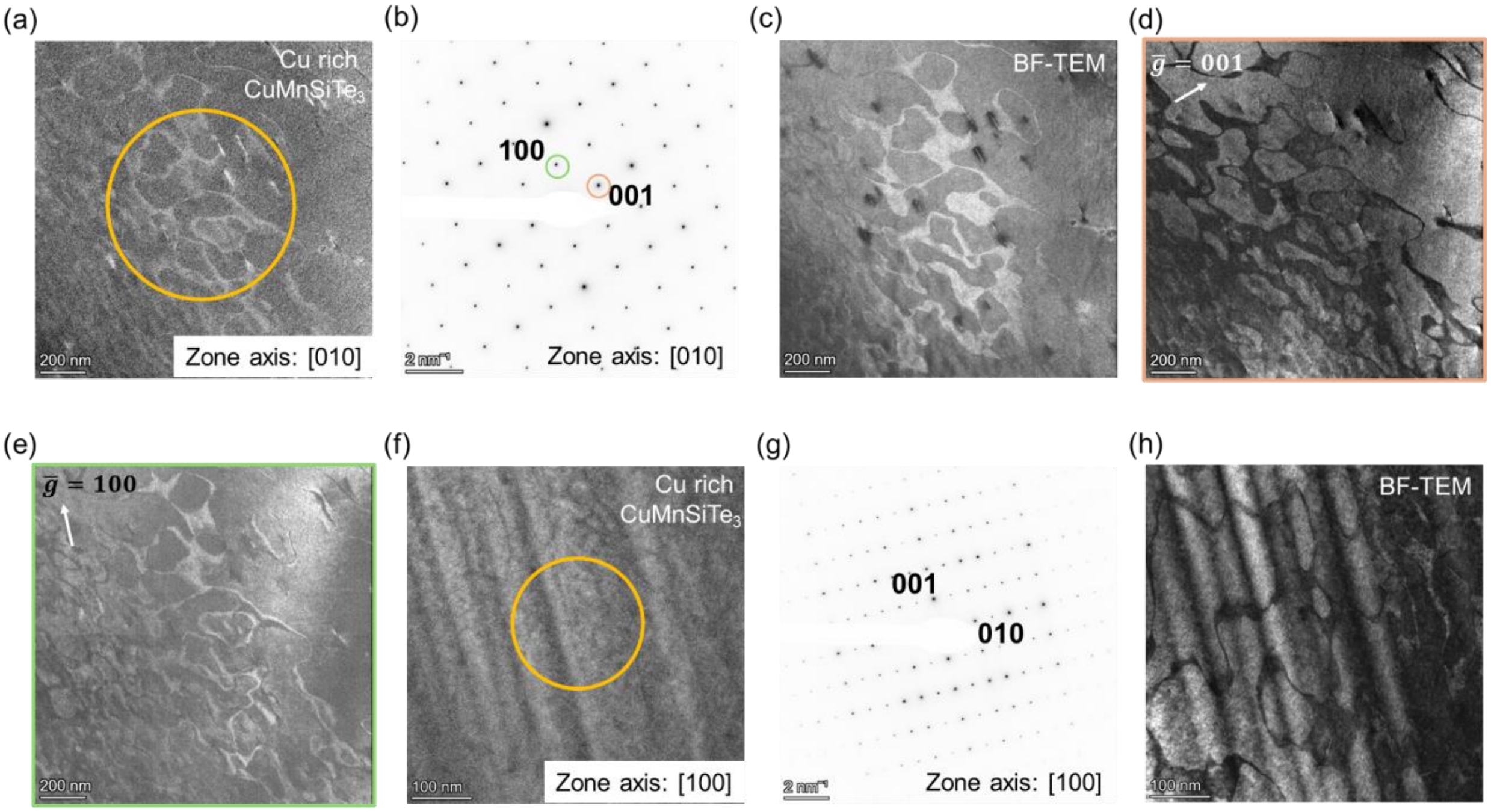

**Figure S19. Loop-like feature investigation:** (a) TEM image of $Cu_{1.14}Mn_{0.77}SiTe_{2.9}$ cross-section specimen taken along the [010] zone axis. (b) SAED pattern from the yellow circled region in (a). The inset shows the reflections used to generate DF-TEM images in (d) and (e). (c) BF-TEM image corresponding to (a). (f) TEM image of $Cu_{1.14}Mn_{0.77}SiTe_{2.9}$ cross-section specimen taken along the [100] zone axis. (g) SAED pattern from the yellow circled region in (f). (h) BF-TEM image corresponding to (f).

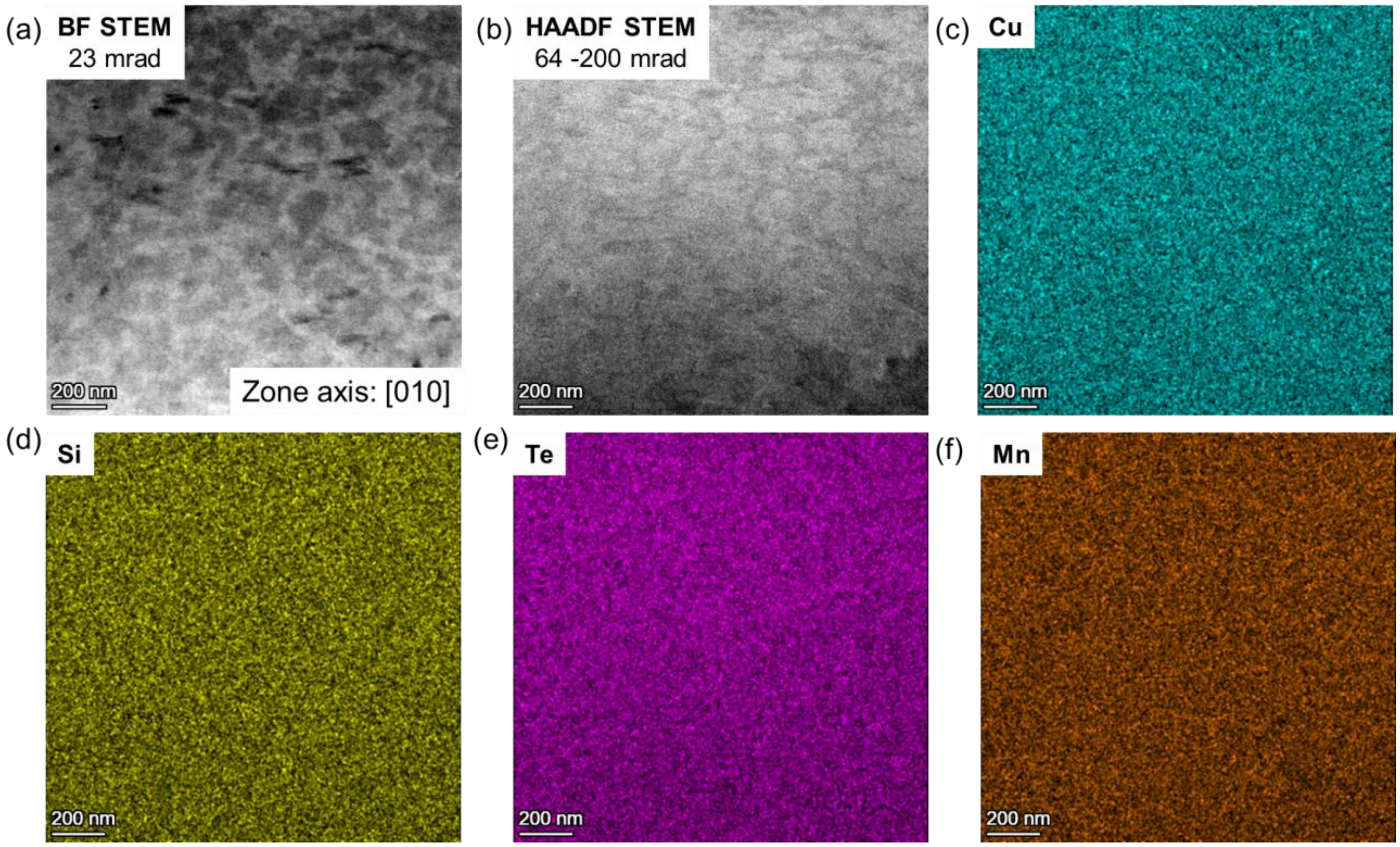

**Figure S20. STEM imaging and elemental distribution maps of loop-like features:** (a) BF-STEM image of $Cu_{1.14}Mn_{0.77}SiTe_{2.9}$ cross-section specimen taken along the [010] zone axis. (b) Corresponding HAADF-STEM image. (c-f) EDS maps of Cu, Si, Te, and Mn, respectively.

**Figure S21: Cu valence state measurement using X-ray photoelectron spectroscopy (XPS) in Cu-rich crystals**

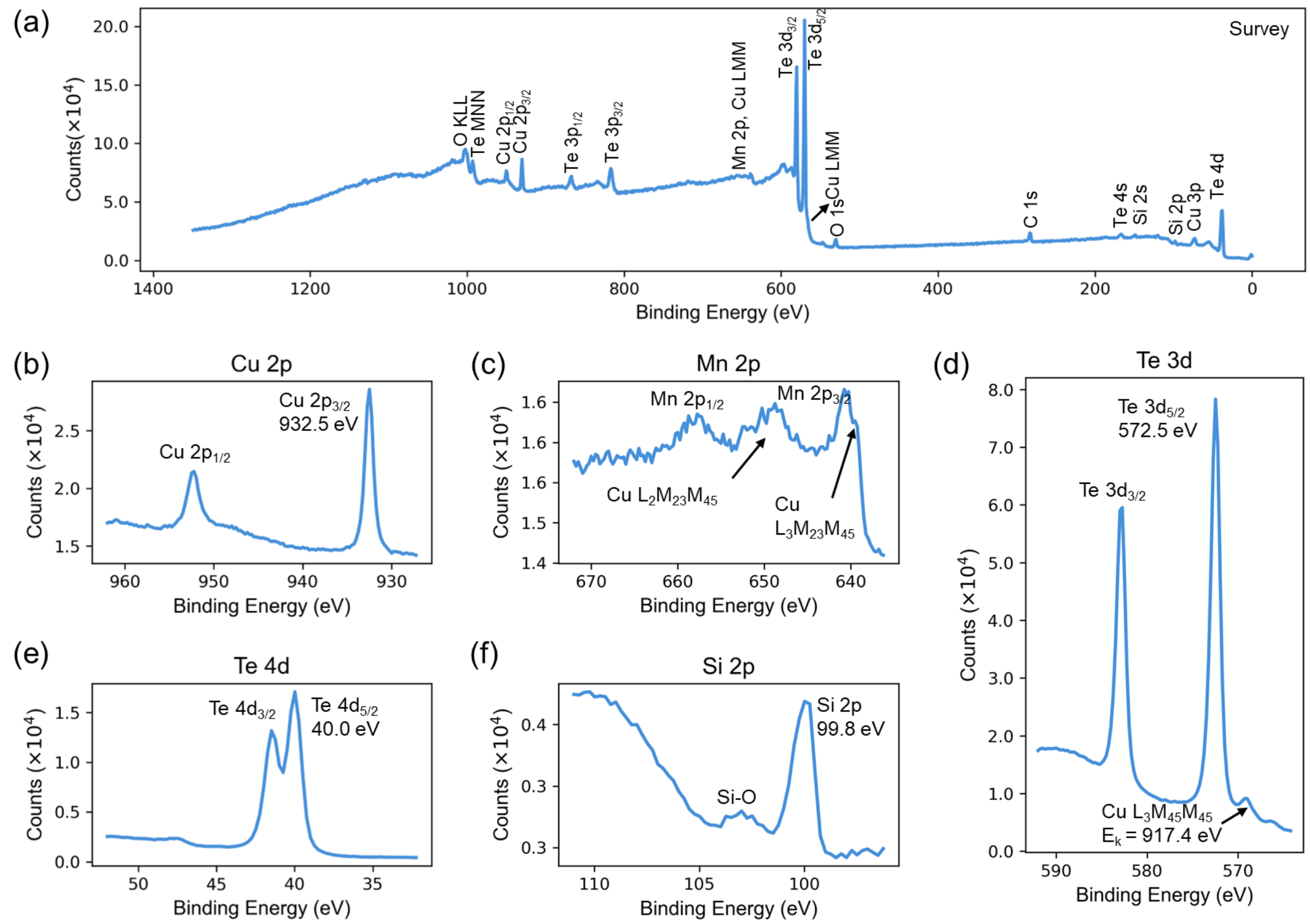

**Figure S21. XPS spectra of Cu-rich crystal.** (a) XPS survey scan, (b) Cu 2p, (c) Mn 2p, (d) Te 3d core level and Cu $L_3M_{45}M_{45}$ Auger spectra, (e)Te 4d, and (f) Si 2p for the Cu-rich sample.

Cu rich bulk crystal sample was cleaved in Argon glove box (<0.5 ppm $H_2O$ and $O_2$) at Penn State followed by inert transfer to XPS chamber. The freshly cleaved side was used to perform the XPS analysis to determine the cation valence in these bulk Cu-rich crystals. Although these samples were cleaved in a glove box, we still see a minor presence of O and C signals in our survey scan.

We use a similar approach of modified Auger parameter (MAP) as our Cu deficient XPS analysis[2] to determine the Cu valence, as this analysis allows us to determine the valence independent of X-ray energy used. Here, $\alpha = E_k (CuL_3M_{45}M_{45}) + E_b (Cu\ 2p_{3/2}) = 917.4$ eV+932.5 eV = 1849.9 eV. The MAP value for many Cu(I) compounds ranges from 1848 to 1850 eV, whereas Cu(0) is consistently ~1851.2 to 1851.3 eV and Cu(II) is ~1850.9 to 1851.7 eV [3]. This indicates that Cu is 1+ in Cu rich sample, similar to Cu deficient sample [2]. Additionally, determining the Mn valence from Mn 2p spectra is challenging due to significant overlap with Cu Auger peaks.